\documentclass[10pt,twocolumn,letterpaper]{article}

\usepackage{iccv}
\usepackage{times}
\usepackage{epsfig}
\usepackage{graphicx}
\usepackage[font={small}]{caption}
\usepackage[font={small}]{subcaption}
\usepackage{amsmath}
\usepackage{amssymb}
\usepackage{float}
\usepackage{enumitem}
\usepackage{xcolor}
\usepackage{multicol}

\usepackage{soul}
\setul{1pt}{.4pt}

\usepackage{etoolbox,refcount}

\newcounter{countitems}
\newcounter{nextitemizecount}
\newcommand{\setupcountitems}{%
  \stepcounter{nextitemizecount}%
  \setcounter{countitems}{0}%
  \preto\item{\stepcounter{countitems}}%
}
\makeatletter
\newcommand{\computecountitems}{%
  \edef\@currentlabel{\number\c@countitems}%
  \label{countitems@\number\numexpr\value{nextitemizecount}-1\relax}%
}
\newcommand{\nextitemizecount}{%
  \getrefnumber{countitems@\number\c@nextitemizecount}%
}
\makeatother
\AtBeginEnvironment{enumerate}{\setupcountitems}
\AtEndEnvironment{enumerate}{\computecountitems}

\makeatletter
\newcommand{\manuallabel}[2]{\def\@currentlabel{#2}\label{#1}}
\makeatother

\usepackage{array}

\newcolumntype{P}[1]{>{\centering\arraybackslash}p{#1}}

\usepackage[pagebackref=true,breaklinks=true,letterpaper=true,colorlinks,bookmarks=false]{hyperref}

\iccvfinalcopy 

\def\iccvPaperID{184} 

\def\figurename{Fig.}
\def\sectionname{Sec.}
\def\tablename{Tab.}
\newenvironment{jlblue}{\color{blue}}{}

\newcommand{\jlreplace}[2]{{\jlblue #2}}

\ificcvfinal\fi
\begin{document}

\title{
Learning Fixed Points in Generative Adversarial Networks: \\ From Image-to-Image Translation to Disease Detection and Localization\vspace{-15pt}}

\author{Md Mahfuzur Rahman Siddiquee\textsuperscript{1}, Zongwei Zhou\textsuperscript{1,3}, Nima Tajbakhsh\textsuperscript{1}, Ruibin Feng\textsuperscript{1},\\
Michael B. Gotway\textsuperscript{2}, Yoshua Bengio\textsuperscript{3}, and Jianming Liang\textsuperscript{1,3}\\
\textsuperscript{1}Arizona State University; \textsuperscript{2}Mayo Clinic; \textsuperscript{3}Mila -- Quebec Artificial Intelligence Institute\\
\vspace{-22pt}}


\twocolumn[{%
\renewcommand\twocolumn[1][]{#1}%
\maketitle
\vspace{-170pt}
\begin{center}
    \centering
    \textcolor{blue}{Please cite the paper as M.M. Rahman Siddiquee, Z. Zhou, N. Tajbakhsh, R. Feng, M. B. Gotway, Y. Bengio, and J. Liang. Learning Fixed Points in Generative Adversarial Networks: From Image-to-Image Translation to Disease Detection and Localization. International Conference on Computer Vision (ICCV), 2019.}
\end{center}%

\vspace{106pt}
\begin{center}
    \centering
    \includegraphics[width=\linewidth]{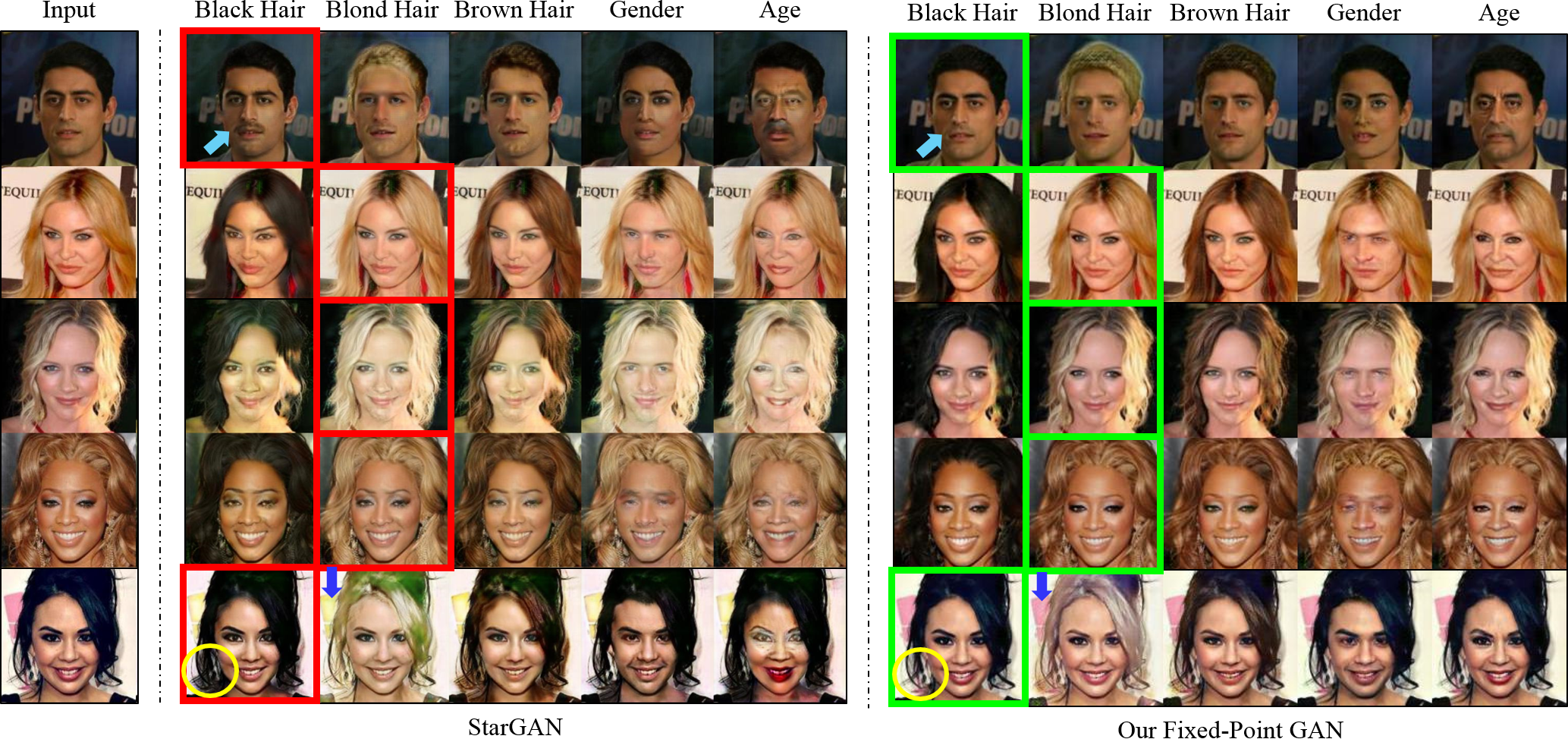}
    \captionof{figure}{[Better viewed on-line in color and zoomed in for details] Comparing our Fixed-Point GAN with StarGAN~\cite{choi2017stargan}, the state of the art in multi-domain image-to-image translation, by translating images into five domains. Combining the domains may yield a same-domain (\eg, black to black hair) or cross-domain (\eg, black to blond hair) translation. For clarity, same-domain translations are framed in red for StarGAN and in green for Fixed-Point GAN. As illustrated, during cross-domain translations, and especially during same-domain translations, StarGAN generates artifacts: introducing a mustache (Row 1, Col.~2; light blue arrow), changing the face colors (Rows 2--5, Cols.~2--6), adding more hair (Row 5, Col.~2; yellow circle), and altering the background (Row 5, Col.~3; blue arrow). Our Fixed-Point GAN overcomes these drawbacks via fixed-point translation learning (see~\sectionname~\ref{sec:method}) and provides a framework for disease detection and localization with only image-level annotation (see~\figurename~\ref{fig:detection_localization_results}).}
    \label{fig:celeba5_stargan_vs_ours}
    \vspace{6pt}
\end{center}%
}]



\begin{figure*}
    \centering
    \includegraphics[width=\linewidth]{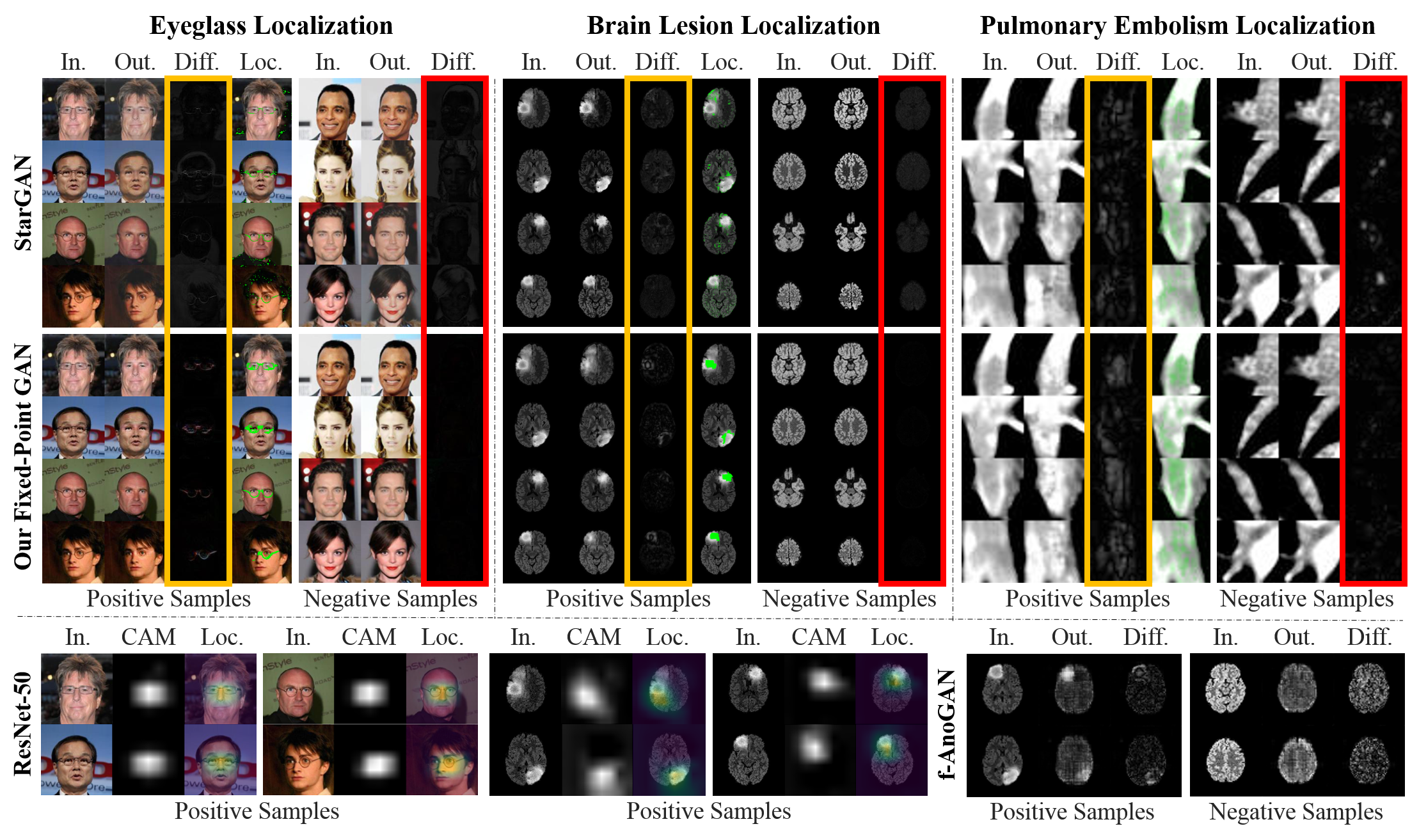}
    \caption{[Better viewed on-line in color and zoomed in for details] Comparing Fixed-Point GAN with the state-of-the-art image-to-image translation~\cite{choi2017stargan}, weakly-supervised localization~\cite{zhou2016learning}, and anomaly detection~\cite{schlegl2019f} for detecting and localizing eyeglasses and diseases using only image-level annotation. Using disease detection as an example, our approach is to translate any image, diseased or healthy, into a healthy image, allowing diseased regions to be revealed by subtracting those two images. Through fixed-point translation learning, our Fixed-Point GAN aims to preserve healthy images during the translation, thereby few differences between the generated (healthy) images and the original (healthy) images are observed in the difference maps (columns framed in red). For diseased images, owing to the transformation learning from diseased images to healthy ones, disease locations are revealed in the difference maps (columns framed in yellow).  For comparison, the localized diseased regions are superimposed on the original images (Loc. Columns), showing that Fixed-Point GAN is more precise than CAM~\cite{zhou2016learning} and f-AnoGAN~\cite{schlegl2019f} for localizing eyeglasses and diseases (bottom row; detailed in \sectionname~\ref{sec:applications}).}
    \label{fig:detection_localization_results}
    \vspace{-6pt}
\end{figure*}


\begin{abstract}

\noindent Generative adversarial networks (GANs) have ushered in a revolution in image-to-image translation. The development and proliferation of GANs raises an interesting question:~can we train a GAN to remove an object, if present, from an image while otherwise preserving the image? Specifically, can a GAN ``virtually heal'' anyone by turning his medical image, with an unknown health status (diseased or healthy), into a healthy one, so that diseased regions could be revealed by subtracting those two images? Such a task requires a GAN to identify a minimal subset of target pixels for domain translation, an ability that we call fixed-point translation, which no GAN is equipped with yet. Therefore, we propose a new GAN, called Fixed-Point GAN, trained by (1) supervising same-domain translation through a conditional identity loss, and (2) regularizing cross-domain translation through revised adversarial, domain classification, and cycle consistency loss. Based on fixed-point translation, we further derive a novel framework for disease detection and localization using only image-level annotation. Qualitative and quantitative evaluations demonstrate that the proposed method outperforms the state of the art in multi-domain image-to-image translation and that it surpasses predominant weakly-supervised localization methods in both disease detection and localization. Implementation is available at \url{https://github.com/jlianglab/Fixed-Point-GAN}.
\end{abstract}
\vspace{-15pt}

\section{Introduction}
\label{sec:introduction}
\noindent Generative adversarial networks (GANs)~\cite{goodfellow2014generative} have proven to be powerful for image-to-image translation, such as changing the hair color, facial expression, and makeup of a person~\cite{choi2017stargan, chang2018pairedcyclegan}, and converting MRI scans to CT scans for radiotherapy planning~\cite{wolterink2017deep}.  Now, the development and proliferation of GANs raises an interesting question: {\em Can GANs remove an object, if present, from an image while otherwise preserving the image content?}  Specifically, can we train a GAN to remove eyeglasses from any image of a face with eyeglasses while keeping unchanged those without eyeglasses? Or, can a GAN ``heal'' a patient on his medical image virtually\footnote{\label{virtalhealing}Virtual healing (see \figurename~\ref{fig:joke} in Appendix) turns an image (diseased or healthy) into a healthy image, thereby subtracting the two images reveals diseased regions.}?
Such a task appears simple, but it actually demands the following four stringent requirements: \\[-17pt]

\begin{itemize}[leftmargin=0.15in]
    \setlength\itemsep{-0.3em}
    \item {\bf Req.~1:} \manuallabel{req1}{1} The GAN must handle unpaired images. It may be too arduous to collect a perfect pair of photos of the same person with and without eyeglasses, and it would be too late to acquire a healthy image for a patient with an illness undergoing medical imaging. 
 
    \item {\bf Req.~2:} \manuallabel{req2}{2} The GAN must require no source domain label when translating an image into a target domain (\ie, source-domain-independent translation). For instance, a GAN trained for virtual healing aims to turn any image, with unknown health status, into a healthy one.
    
     \item {\bf Req.~3:}\manuallabel{req3}{3}  The GAN must conduct an identity transformation for same-domain translation. For ``virtual healing'', the GAN should leave a healthy image intact, injecting neither artifacts nor new information into the image.

     \item {\bf Req.~4:}\manuallabel{req4}{4}  The GAN must perform a minimal image transformation for cross-domain translation. Changes should be applied only to the image attributes directly relevant to the translation task, with no impact on unrelated attributes. For instance, removing eyeglasses should not affect the remainder of the image (\eg, the hair, face color, and background), or removing diseases from a diseased image should not impact the region of the image labeled as normal. 
\end{itemize}

\noindent Currently, no single image-to-image translation method satisfies all aforementioned requirements. The conventional GANs for image-to-image translation~\cite{isola2016image}, although successful, require paired images. CycleGAN~\cite{zhu2017unpaired} mitigates this limitation through cycle consistency, but it still requires two dedicated generators for each pair of image domains resulting a scalability issue due to a requirement for dedicated generators. CycleGAN also fails to support source-domain-independent translation: selecting the suitable generator requires labels for both the source and target domain. StarGAN~\cite{choi2017stargan} overcomes both limitations by learning one single  generator for all domain pairs of interest.  
However, StarGAN has its own shortcomings. First, StarGAN tends to make unnecessary changes during cross-domain translation. As illustrated in \figurename~\ref{fig:celeba5_stargan_vs_ours}, StarGAN tends to alter the face color, although the goal of domain translation is to change the gender, age, or hair color in images from the CelebFaces dataset~\cite{liu2015deep}. Second, StarGAN fails to competently handle same-domain translation. Referring to examples framed with red boxes in \figurename~\ref{fig:celeba5_stargan_vs_ours}, StarGAN needlessly adds a mustache to the face in Row 1, and unnecessarily alters the hair color in Rows 2--5, where only a simple identity transformation is desired. These shortcomings may be acceptable for image-to-image translation in natural images, but in sensitive domains, such as medical imaging, they may lead to dire consequences---unnecessary changes and artifacts introduction may result in misdiagnosis. Furthermore, overcoming the above limitations is essential for adapting GANs for object/disease detection, localization, segmentation---and removal.

\begin{figure*}[t]
    \centering 
    \includegraphics[width=1.0\textwidth]{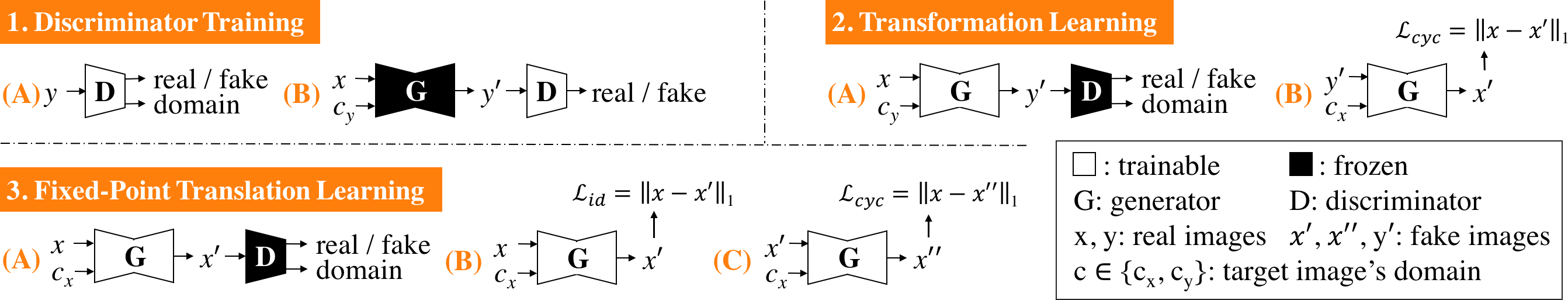}
    \vspace{-3pt}
    \caption{Fixed-Point GAN training scheme. Similar to StarGAN, our discriminator learns to distinguish real/fake images and classify the domains of input images (1A--B). However, unlike StarGAN, our generator learns to perform not only cross-domain translations via transformation learning (2A--B), but also same-domain translations via fixed-point translation learning (3A--C), which is essential for mitigating the limitations of StarGAN (\figurename~\ref{fig:celeba5_stargan_vs_ours}) and realizing disease detection and localization using only image-level annotation (\figurename~\ref{fig:detection_localization_results}).}
    \label{fig:method}
    \vspace{-6pt}
\end{figure*}

Therefore, we propose a novel GAN. We call it Fixed-Point GAN for its new fixed-point\footnote{Mathematically, $x$ is a fixed point of function $f(\cdot)$ if $f(x) = x$. We borrow the term to describe the pixels to be preserved when applying the GAN translation function.} translation ability, which allows the GAN to identify a minimal subset of pixels for domain translation. 
To achieve this capability, we have devised a new training scheme to promote the fixed-point translation during training (\figurename~\ref{fig:method}-3) by (1) supervising same-domain translation through an additional conditional identity loss (\figurename~\ref{fig:method}-3B), and (2) regularizing cross-domain translation through revised adversarial (\figurename~\ref{fig:method}-3A), domain classification (\figurename~\ref{fig:method}-3A), and cycle consistency (\figurename~\ref{fig:method}-3C) loss. Owing to its fixed-point translation ability, Fixed-Point GAN performs a minimal transformation for cross-domain translation and strives for an identity transformation for same-domain translation. Consequently, Fixed-Point GAN not only achieves better image-to-image translation for natural images but also offers a novel framework for disease detection and localization with only image-level annotation. Our experiments demonstrate that Fixed-Point GAN significantly outperforms StarGAN over multiple datasets for the tasks of image-to-image translation and predominant anomaly detection and weakly-supervised localization methods for disease detection and localization. 
Formally, we make the following contributions:
\begin{enumerate}
\setlength\itemsep{-0.3em}

    \item We introduce a new concept: fixed-point translation, leading to a new GAN: Fixed-Point GAN.
    
    \item We devise a new scheme to train fixed-point translation by supervising same-domain translation and regularizing cross-domain translation.

   \item We show that Fixed-Point GAN outperforms the state-of-the-art method in image-to-image translation for both natural and medical images.
   
   \item We derive a novel method for disease detection and localization using image-level annotation based on fixed-point translation learning.

    \item We demonstrate that our disease detection and localization method based on Fixed-Point GAN is superior to not only its counterpart based on the state-of-the-art image-to-image translation method but also superior to predominant weakly-supervised localization and anomaly detection methods.
\end{enumerate}
\noindent Our Fixed-Point GAN has the potential to exert important clinical impact on computer-aided diagnosis in medical imaging, because it requires only image-level annotation for training. Obtaining image-level annotation is far more feasible and practical than manual lesion-level annotation, as a large number of diseased and healthy images can be collected from the picture archiving and communication systems, and labeled at the image level by analyzing their radiological reports with NLP. With the availability of large databases of medical images and their corresponding radiological reports, we envision not only that Fixed-Point GAN will detect and localize diseases more accurately, but also that it may eventually be able to ``cure''\textsuperscript{\ref{virtalhealing}}, thus segment diseases in the future.

\section{Related Work}
\label{sec:related}
Fixed-Point GAN can be used for image-to-image translation as well as disease detection and localization with only image-level annotation. Hence, we first compare our Fixed-Point GAN with other image-to-image translation methods, and then explain how Fixed-Point GAN differs from the  weakly-supervised lesion localization and anomaly detection methods suggested in medical imaging.

\label{subsec:related_img2img}

\noindent {\bf Image-to-image translation:} The literature surrounding GANs~\cite{goodfellow2014generative} for image-to-image translation is extensive~\cite{isola2016image, zhu2017unpaired, kim2017learning, zhu2017toward, liu2017unsupervised, yi2017dualgan, choi2017stargan, ledig2017photo}; therefore we limit our discussion to only the most relevant works. CycleGAN~\cite{zhu2017unpaired} has made a breakthrough in {\em unpaired} image-to-image translation via cycle consistency. Cycle consistency has proven to be effective in preserving object shapes in translated images, but it may not preserve other image attributes, such as color; therefore, when converting Monet's painting to photos (a cross-domain translation), Zhu \etal~\cite{zhu2017unpaired} imposes an extra identity loss to preserve the colors of input images. However, identity loss cannot be used for cross-domain translation in general, as it would limit the transformation power. For instance, it would make it impossible to translate black hair to blond hair. Therefore, unlike CycleGAN, we conditionally incorporate the identity loss only during fixed-point translation learning for same-domain translations. Moreover, during inference, CycleGAN requires that the source domain be provided, thereby violating our Req.~\ref{req2} as discussed in \sectionname~\ref{sec:introduction} and rendering CycleGAN unsuitable for our purpose. 
StarGAN~\cite{choi2017stargan} empowers a single generator with the capability for {\em multi-domain} image-to-image translation, and does not require the source domain of the input image at inference time. 
However, StarGAN has its own shortcomings, which violate Reqs.~\ref{req3} and \ref{req4} as discussed in \sectionname~\ref{sec:introduction}. Our Fixed-Point GAN overcomes StarGAN's shortcomings, not only dramatically improving image-to-image translation but also opening the door to an innovative use of the generator as a disease detector and localizer (Figs.\ref{fig:celeba5_stargan_vs_ours}-\ref{fig:detection_localization_results}). 

\label{subsec:related_weakly}
\noindent{\bf Weakly-supervised localization:} Our work is also closely related to weakly-supervised localization, which, in natural imaging, is commonly tackled by  saliency map~\cite{simonyan2013deep}, global max pooling~\cite{oquab2015object}, and class activation map (CAM) based on global average pooling (GAP)~\cite{zhou2016learning}. In particular, the CAM technique has recently been the subject of further research, resulting in several extensions with improved localization power. Pinheiro and Collobert~\cite{pinheiro2015image} replaced the original GAP with a log-sum-exponential pooling layer, while other works~\cite{singh2017hide,zhang2018adversarial} aim to force the CAM to discover the complementary parts rather than just the most discriminative parts of the objects. Selvaraju et al.~\cite{selvaraju2017grad} proposed GradCAM where the weights used to generate the CAM come from gradient backpropagation; that is, the weights depend on the input image as opposed to the fixed pre-trained weights used in the original CAM.  

Despite the extensive literature in natural imaging, weakly supervised localization in medical imaging has taken off only recently. Wang et al.~\cite{wang2017chestx} used the CAM technique for the first time for lesion localization in chest X-rays. The following research works, however, either combined the original CAM with extra information (\eg, limited fine-grained annotation \cite{li2018thoracic,shin2018joint,bai2017semi} and disease severity-level \cite{tang2018attention}), or slightly extended the original CAM with no significant localization gain. Noteworthy, as evidenced by~\cite{cai2018iterative}, the adoption of more advanced versions of the CAM such as the complementary-discovery algorithm \cite{singh2017hide,zhang2018adversarial} has not proved promising for weakly-supervised lesion localization in medical imaging.  Different from the previous works, Baumgartner \etal ~\cite{baumgartner2017visual} propose VA-GAN to learn the difference between a healthy brain and the one affected by Alzheimer's disease. Although unpaired, VA-GAN requires that all images be registered; otherwise, it fails to preserve the normal brain structures (see the appendix for illustrations). Furthermore, VA-GAN requires the source-domain label at inference time (input image being healthy or diseased), thus violating our Req.~\ref{req2} as listed in \sectionname~\ref{sec:introduction}. Therefore, the vanilla CAM remains as a strong performance baseline for weakly-supervised lesion localization in medical imaging.

 To our knowledge, we are among the first to develop GANs based on image-to-image translation for disease detection and localization with image-level annotation only. Both qualitative and quantitative results suggest that our image-translation-based approach provides more precise localization than the CAM-based method~\cite{zhou2016learning}.

\label{subsec:related_detection}

\noindent {\bf Anomaly detection:} Our work may seem related to anomaly detection  \cite{chen2018unsupervised,schlegl2019f,alex2017generative} where the task is to detect {\em rare} diseases by learning from only {\em healthy} images. 
Chen \etal~\cite{chen2018unsupervised} use an adversarial autoencoder to learn healthy data distribution. The anomalies are identified by feeding a diseased image to the trained autoencoder followed by subtracting the reconstructed diseased image from the input diseased image. The method suggested by
Schlegl \etal~\cite{schlegl2019f} learns a generative model of healthy training data through a GAN, which receives a random latent vector as input and then attempts to distinguish between real and generated fake healthy images. They further propose a fast mapping  that can identify anomalies of the diseased images by projecting the diseased data into the GAN's latent space. Similar to \cite{schlegl2019f}, Alex \etal~\cite{alex2017generative} use a GAN to learn a generative model of healthy data.
To identify anomalies, they scan an image pixel-by-pixel and feed the scanned crops to the discriminator of the trained GAN. An anomaly map is then constructed by putting together the anomaly scores by the discriminator.

However,  Fixed-Point GAN is different from anomaly detectors in both training and functionality.

Trained using only the healthy images, anomaly detectors cannot distinguish between different types of anomalies, as they treat all anomalies as ``a single category''. In contrast, our Fixed-Point GAN can take advantage of anomaly labels, if available, enabling both localization and recognition of all anomalies. Nevertheless, for a comprehensive analysis, we have compared Fixed-Point GAN against \cite{schlegl2019f} and \cite{alex2017generative}.

\section{Method}
\label{sec:method}

In the following, we present a high-level overview of Fixed-Point GAN, followed by a detailed mathematical description of each individual loss function.

Like StarGAN, our discriminator is trained to classify an image as real/fake and its associated domain (\figurename~\ref{fig:method}-1). Using our new training scheme, the generator learns both cross- and same-domain translation, which differs from StarGAN, wherein the generator only learns the former. Mathematically, for any input $x$ from domain $c_x$ and target domain $c_y$, the StarGAN generator learns to perform cross-domain translation ($c_x \neq c_y$), $G(x, c_y) \xrightarrow[]{} y'$,  where $y'$ is the image in domain $c_y$. Since $c_y$ is selected randomly during training of StarGAN, there is a slender chance that $c_y$ and $c_x$ turn out identical, but StarGAN is not designed to learn same-domain translation explicitly. The Fixed-Point GAN generator, in addition to learning the cross-domain translation, learns to perform the same-domain translation as 
    $G(x, c_x) \xrightarrow[]{} x'$.

Our new fixed-point translation learning (\figurename~\ref{fig:method}-3) not only enables same-domain translation but also regularizes cross-domain translation (\figurename~\ref{fig:method}-2) by encouraging the generator to find a minimal transformation function, thereby penalizing changes unrelated to the present domain translation task. Trained for only cross-domain image translation, StarGAN cannot benefit from such regularization, resulting in many artifacts as illustrated in \figurename~\ref{fig:celeba5_stargan_vs_ours}. Consequently, our new training scheme offers three advantages: (1) reinforced same-domain translation, (2) regularized cross-domain translation, and (3) source-domain-independent translation. To realize these advantages, we define the loss functions of Fixed-Point GAN as follows: 

\noindent\textbf{Adversarial Loss.} In the proposed method, the generator learns the cross- and same-domain translations. To ensure the generated images appear realistic in both scenarios, the adversarial loss is revised as follows and the modification is highlighted in \tablename~\ref{tab:loss_comparison}:
\begin{equation}
\label{eq:adv_loss}
\begin{split}
    \mathcal{L}_{adv} = & \sum_{c \in \{c_x, c_y\}} \mathbb{E}_{x, c}[\log{(1 - D_{real/fake}(G(x, c)))}] \\
    & + \mathbb{E}_{x}[\log{D_{real/fake}(x)}]
\end{split}
\end{equation}

\noindent\textbf{Domain Classification Loss.}  The adversarial loss ensures the generated images appear realistic, but it cannot guarantee domain correctness. As a result, the discriminator is trained with an additional domain classification loss, which forces the generated images to be of the correct domain. The domain classification loss for the discriminator is identical to that of StarGAN,
\begin{equation}
\label{eq:real_domain_loss}
    \mathcal{L}_{domain}^{r} = \mathbb{E}_{x, c_x}[-\log{D_{domain}(c_x|x)}]
\end{equation}
\noindent but we have updated the domain classification loss for the generator to account for both same- and cross-domain translations, ensuring that the generated image is from the correct domain in both scenarios: 
\begin{equation}
\label{eq:fake_domain_loss}
    \mathcal{L}_{domain}^{f} = \sum_{c \in \{c_x, c_y\}}\mathbb{E}_{x, c}[-\log{D_{domain}(c|G(x, c))}]
\end{equation}

\noindent\textbf{Cycle Consistency Loss.} Optimizing the generator, for unpaired images, with only the adversarial loss has multiple possible, but random, solutions. The additional \textit{cycle consistency loss} (Eq.~\ref{eq:cyc_loss}) helps the generator to learn a transformation that can preserve enough input information, such that the generated image can be translated back to original domain. Our modified \textit{cycle consistency loss} ensures that both cross- and same-domain translations are cycle consistent.
\begin{equation}
\label{eq:cyc_loss}
\begin{split}
    \mathcal{L}_{cyc} =\ & \mathbb{E}_{x, c_x, c_y}[||G(G(x, c_y), c_x) - x||_1] \ + \\
    & \mathbb{E}_{x, c_x}[||G(G(x, c_x), c_x) - x||_1]
\end{split}
\end{equation}

\noindent\textbf{Conditional Identity Loss.} During training, StarGAN~\cite{choi2017stargan} focuses on translating the input image to different target domains. This strategy cannot penalize the generator when it changes aspects of the input that are irrelevant to match target domains (\figurename~\ref{fig:celeba5_stargan_vs_ours}). In addition to learning a translation to different domains, we force the generator, using the \textit{conditional identity loss} (Eq.~\ref{eq:conditional_id_loss}), to preserve the domain identity while translating the image to the source domain. This also helps the generator learn a minimal transformation for translating the input image to the target domain.
\begin{equation}
\label{eq:conditional_id_loss}
    \mathcal{L}_{id} = \begin{cases}
      0,  & c=c_y  \\
      \mathbb{E}_{x, c}[||G(x, c) - x||_1], & c=c_x  \\
    \end{cases}
\end{equation}

\begin{table}[t]
    \begin{center}
    \includegraphics[width=\linewidth]{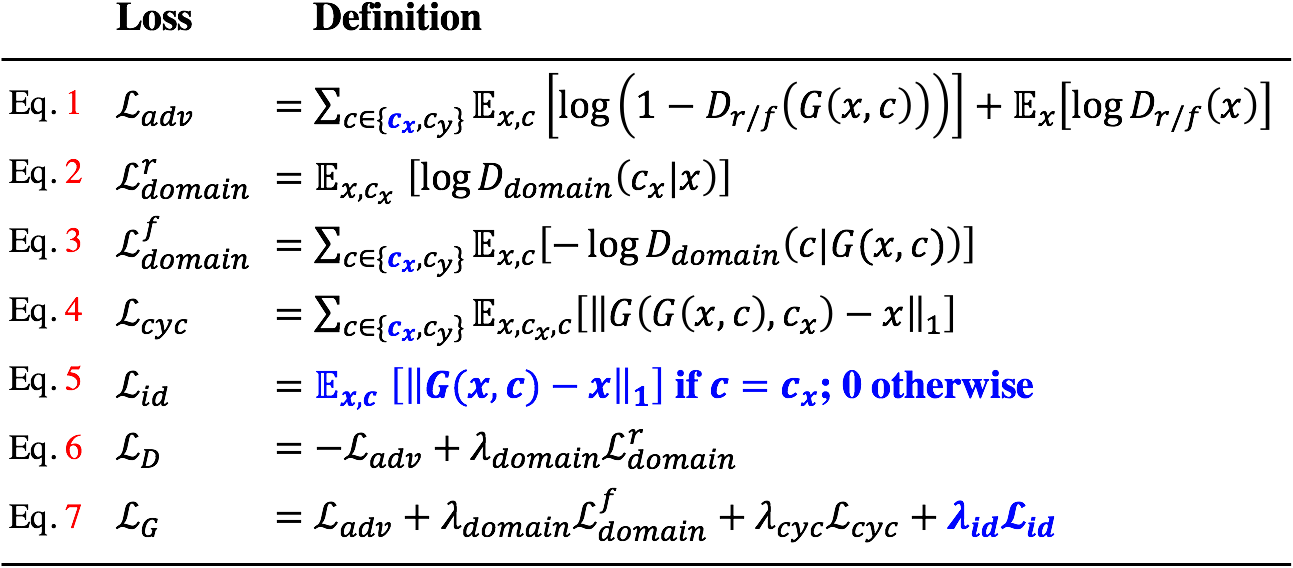}
    \end{center}

    \caption{
    Loss functions in Fixed-Point GAN. Terms inherited from StarGAN are in black, while highlighted in {\color{blue} blue} are our modifications to mitigate StarGAN's limitations (\figurename~\ref{fig:celeba5_stargan_vs_ours}).
    }
    \label{tab:loss_comparison}
    \vspace{-0.1in}
\end{table}

\noindent\textbf{Full Objective.} Combining all losses, the final full objective function for the discriminator and generator can be described by Eq.~\ref{eq:d_full_loss} and Eq.~\ref{eq:g_full_loss}, respectively.
\begin{equation}
\label{eq:d_full_loss}
    \mathcal{L}_D = -\mathcal{L}_{adv} + \lambda_{domain}\mathcal{L}_{domain}^{r}
\end{equation}
\noindent
\begin{equation}
\label{eq:g_full_loss}
    \mathcal{L}_G = \mathcal{L}_{adv} + \lambda_{domain}\mathcal{L}_{domain}^{f} + \lambda_{cyc}\mathcal{L}_{cyc} +  \lambda_{id}\mathcal{L}_{id}
\end{equation}
where $\lambda_{domain}$, $\lambda_{cyc}$, and $\lambda_{id}$ determine the relative importance of the \textit{domain classification loss}, \textit{cycle consistency loss}, and \textit{conditional identity loss}, respectively. \tablename~\ref{tab:loss_comparison} summarizes the loss functions of Fixed-Point GAN.

\section{Applications}
\label{sec:applications}

\subsection{Multi-Domain Image-to-Image Translation}

\noindent\textbf{Dataset.} To compare the proposed Fixed-Point GAN with StarGAN~\cite{choi2017stargan} (the current state of the art), we use the CelebFaces Attributes (CelebA) dataset~\cite{liu2015deep}. This dataset is composed of a total of 202,599 facial images of various celebrities, each with 40 different attributes. Following StarGAN's public implementation~\cite{choi2017stargan}, we adopt 5 domains (\texttt{black hair}, \texttt{blond hair}, \texttt{brown hair}, \texttt{male}, and \texttt{young}) for our experiments and pre-process the images by cropping the original 178$\times$218 images into 178$\times$178 and then re-scaling to 128$\times$128. We use a random subset of 2,000 samples for testing and the remainder for training.

\noindent\textbf{Method and Evaluation.} We evaluate the cross-domain image translation quantitatively by classification accuracy and qualitatively by changing one attribute (\eg hair color, gender, or age) at a time from the source domain. This step-wise evaluation facilitates tracking  changes to image content. We also evaluate the same-domain image translation both qualitatively and quantitatively by measuring image-level $L_1$ distance between the input and translated images.

\noindent\textbf{Results.} \figurename~\ref{fig:celeba5_stargan_vs_ours} presents a qualitative comparison between StarGAN and Fixed-Point GAN for multi-domain image-to-image translation. For  the cross-domain image translation, StarGAN tends to make unnecessary changes, such as altering the face color when the goal of translation is to change the gender, age, or hair color (Rows 2--5 in \figurename~\ref{fig:celeba5_stargan_vs_ours}). Fixed-Point GAN, however, preserves the face color while successfully translating the images to the target domains. Furthermore, Fixed-Point GAN preserves  the image background (marked with a blue arrow in Row 5 of \figurename~\ref{fig:celeba5_stargan_vs_ours}), but StarGAN fails to do so. This capability of Fixed-Point GAN is further supported by our quantitative results in \tablename~\ref{tab:cross_domain_classification}.

The superiority of Fixed-Point GAN over StarGAN is even more striking for the same-domain image translation. As shown in \figurename~\ref{fig:celeba5_stargan_vs_ours}, Fixed-Point GAN effectively keeps the image content intact (images outlined in green) while StarGAN undesirably changes the image content (images outlined in red). For instance, the input image in the fourth row of \figurename~\ref{fig:celeba5_stargan_vs_ours} is from the domains of blond hair, female, and young. The same domain translation with StarGAN results in an image in which the hair and face colors are significantly altered. Although this color is closer to the average blond hair color in the dataset, it is far from that in the input image. Fixed-Point GAN, with fixed-point translation ability, handles this problem properly. Further qualitative comparisons between StarGAN and Fixed-Point GAN are provided in the appendix.

\tablename~\ref{tab:image_level_l1} presents a quantitative comparison between StarGAN and Fixed-Point GAN for the task of same-domain image translation. We use the image-level $L_1$ distance between the input and generated images as the performance metric. To gain additional insights into the comparison, we have included a dedicated autoencoder model that has the same architecture as the generator used in StarGAN and Fixed-Point GAN. As seen, the dedicated autoencoder has an image-level $L_1$ reconstruction error of 0.11$\pm$0.09, which can be regarded as a technical lower bound for the reconstruction error. Fixed-Point GAN dramatically reduces the reconstruction error of StarGAN from 2.40$\pm$1.24 to 0.36$\pm$0.35. Our quantitative comparisons are commensurate with the qualitative results shown in \figurename~\ref{fig:celeba5_stargan_vs_ours}.

\begin{table}
    \begin{center}
    \setlength\tabcolsep{3.5pt}
    \begin{tabular}{c c c}
        \textbf{Real Images (Acc.)} & \textbf{Our Fixed-Point GAN} & \textbf{StarGAN} \\
        \hline
        $94.5\%$ & $92.31\%$ & $90.82\%$\\
        \hline
    \end{tabular}
    \end{center}

    \caption{Comparison between the quality of images generated by StarGAN and our method. For this purpose, we have trained a classifier on all 40 attributes of CelebA dataset, which achieves $94.5\%$ accuracy on real images, meaning that the generated images should also have the same classification accuracy to look as realistic as the real images. As seen, the quality of generated images by Fixed-Point GAN is closer to real images, underlining the necessity and effectiveness of fixed-point translation learning in cross-domain translation.}
    \label{tab:cross_domain_classification}
\end{table}

\begin{table}
    \begin{center}
    \begin{tabular}{c c c}
        \textbf{Autoencoder} & \textbf{Our Fixed-Point GAN} & \textbf{StarGAN} \\
        \hline
        $0.11~\pm~0.09$ & $0.36~\pm~0.35$ & $2.40~\pm~1.24$\\
        \hline
    \end{tabular}
    \end{center}

    \caption{Image-level $L_1$ distance comparison for same-domain translation. Fixed-Point GAN achieves significantly lower same-domain translation error than StarGAN, approximating the lower bound error that can be achieved by a stand-alone autoencoder.}
    \label{tab:image_level_l1}
    \vspace{-0.1in}
\end{table}

\begin{figure*}[t]
     \centering
     \begin{subfigure}[b]{0.22\linewidth}
         \centering
         \captionsetup{justification=centering}
         \caption{Image-level detection\\on synthetic images}
         \includegraphics[width=\linewidth]{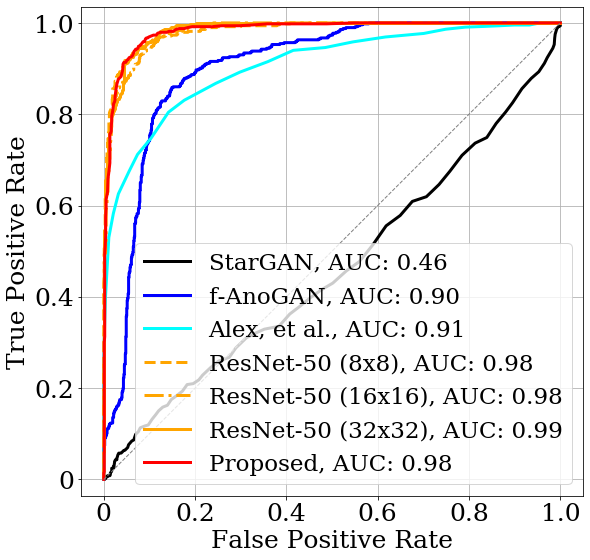}
         \label{subfig:brats_roc}
     \end{subfigure}
     \begin{subfigure}[b]{0.27\linewidth}
         \centering
         \captionsetup{justification=centering}
         \caption{Lesion-level localization\\on synthetic images}
         \includegraphics[width=\linewidth]{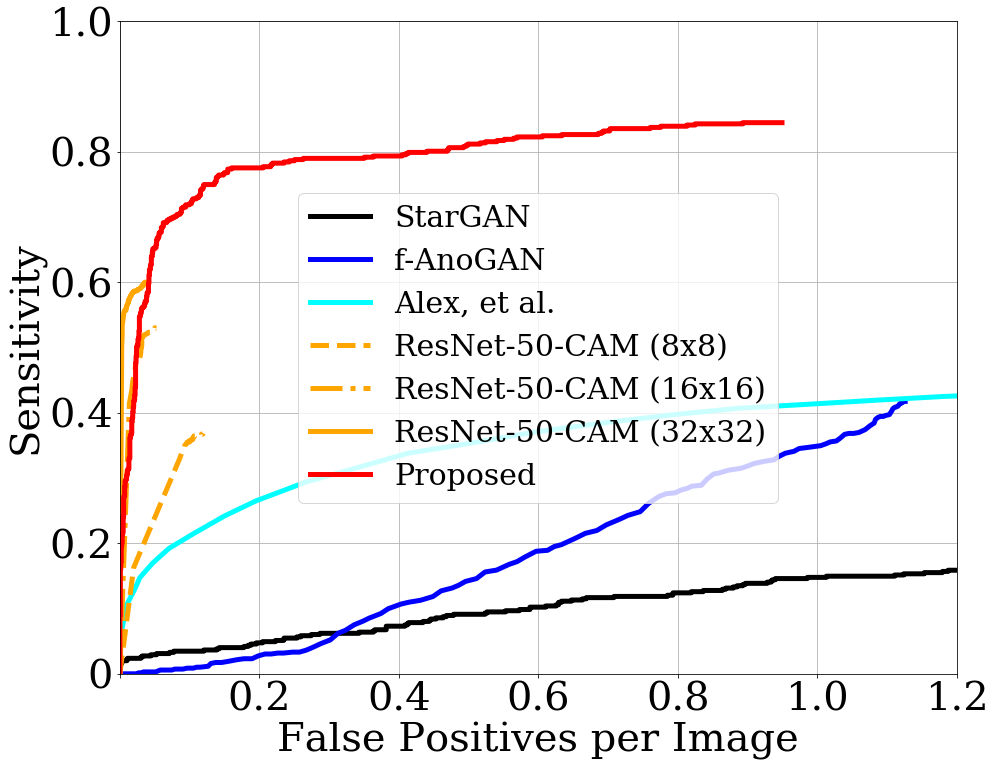}
         \label{subfig:brats_froc}
     \end{subfigure}
     \begin{subfigure}[b]{0.22\linewidth}
         \centering
         \captionsetup{justification=centering}
         \caption{Image-level detection\\on real images}
         \includegraphics[width=\linewidth]{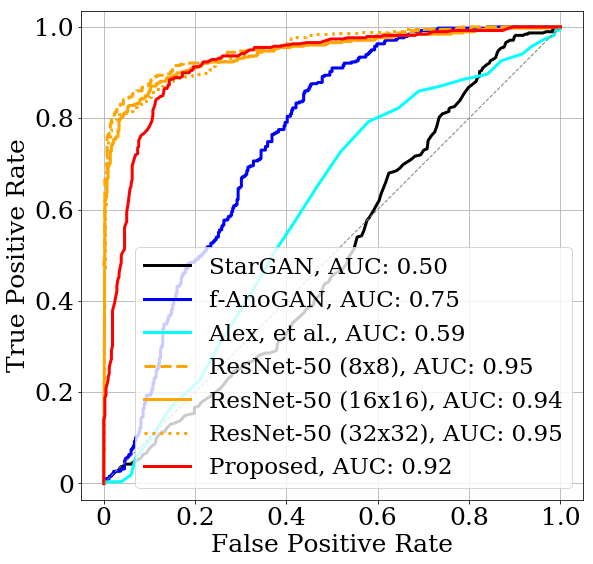}
         \label{subfig:brats_real_roc}
     \end{subfigure}
     \begin{subfigure}[b]{0.27\linewidth}
         \centering
         \captionsetup{justification=centering}
         \caption{Lesion-level localization\\on real images}
         \includegraphics[width=\linewidth]{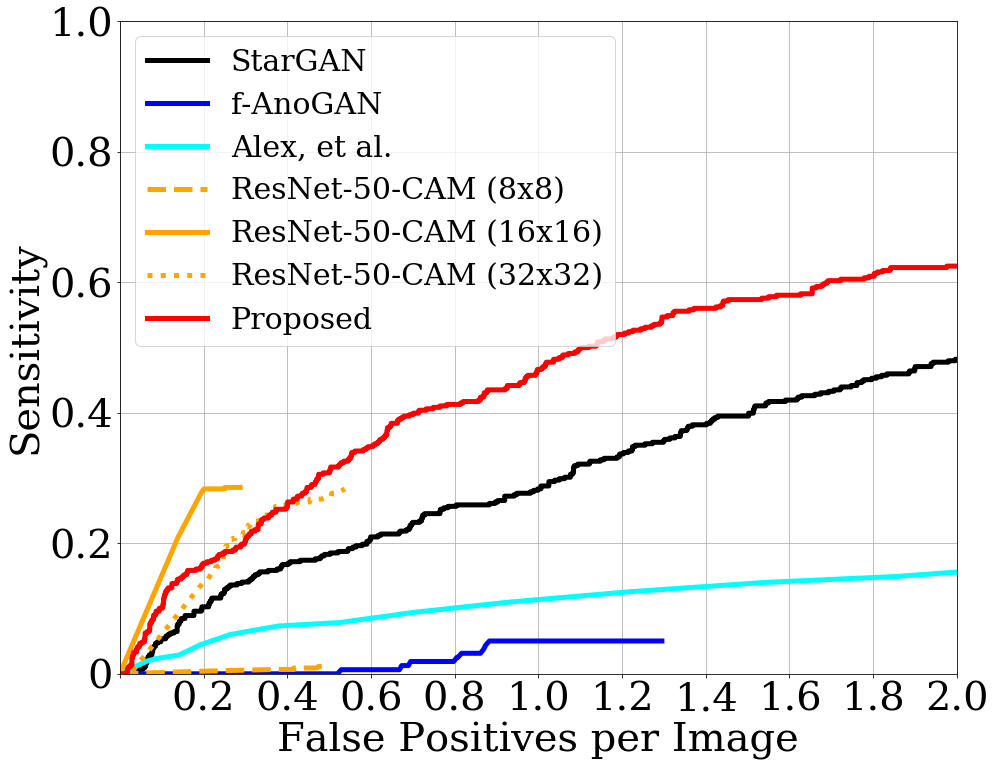}
         \label{subfig:brats_real_froc}
     \end{subfigure}
     \vspace{-8pt}
     \caption{Comparing Fixed-Point GAN with StarGAN, f-AnoGAN, GAN-based brain lesion detection method by Alex, \etal~\cite{alex2017generative}, and ResNet-50 on BRATS 2013. ROCs for image-level detection and FROCs for lesion-level localization on synthetic brain images are provided in (a), (b) respectively and on real brain images in (c), (d) respectively.}
     \label{fig:brats_result}
     \vspace{-6pt}
\end{figure*}

\subsection{Brain Lesion Detection and Localization with Image-Level Annotation}
\label{subsec:application_brain}

\noindent\textbf{Dataset.} We extend Fixed-Point GAN from an image-to-image translation method to a weakly supervised brain lesion detection and localization method, which requires only image-level annotation. As a proof of concept, we use the BRATS 2013 dataset~\cite{menze2015multimodal,info:doi/10.2196/jmir.2930}. BRATS 2013 consists of synthetic and real images. We randomly split the synthetic and real images at the patient-level into 40/10 and 24/6 for training/testing, respectively.  More details about the dataset selection are provided in the appendix.

\noindent\textbf{Method and Evaluation.} For training we use only image-level annotation (healthy/diseased). Fixed-Point GAN is trained for the cross-domain translation (diseased images to healthy images and vice versa) as well as the same-domain translation using the proposed method. At inference time, we focus on translating any images into the healthy domain. The desired GAN behaviour is to translate diseased images to healthy ones while keeping healthy images intact. Having translated the images into the healthy domain, we then detect the presence and location of a lesion in the difference image by subtracting the translated healthy image from the input image. We refer the resultant image as \textit{difference map}. 

We evaluate the difference map at two different levels: (1) image-level disease detection and (2) lesion-level  localization. For image-level detection, we take the maximum value across all pixels in the difference map as the \textit{detection score}. We then use receiver operating characteristics (ROC) analysis for performance evaluation. For the lesion-level localization task, we first binarize the difference maps using color quantization followed by a connected component analysis. Each connected component with an area larger than 10 pixels is considered as a lesion candidate. A lesion  is considered ``detected" if the centroid of at least a lesion candidate falls inside the lesion ground truth.

We evaluate Fixed-Point GAN in comparison with 
StarGAN~\cite{choi2017stargan}, CAM~\cite{zhou2016learning}, f-AnoGAN~\cite{schlegl2019f}, GAN-based brain lesion detection method proposed by Alex, \etal~\cite{alex2017generative}. Comparison with StarGAN allows us to study the effect of the proposed fixed-point translation learning. We choose CAM for comparison because it covers an array of weakly-supervised localization works in medical imaging~\cite{wang2017chestx,tang2018attention,hwang2016self}, and as discussed in \sectionname~\ref{sec:related}, it is arguably a strong performance baseline for comparison. We train a standard ResNet-50 classifier~\cite{he2016deep} and compute CAM following~\cite{zhou2016learning} for localization, referring as ResNet-50-CAM in the rest of this paper. To get higher resolution CAMs, we truncate ResNet-50 at three levels and report localization performance in 8$\times$8, 16$\times$16, and 32$\times$32 feature maps. Although \cite{schlegl2019f} and \cite{alex2017generative} stand as state of the art for anomaly detection, we select them for more comparison since they also fulfill the task requirements. We use the official implementation of \cite{schlegl2019f}.

\noindent\textbf{Results.} \figurename~\ref{subfig:brats_roc} compares the ROC curves of Fixed-Point GAN and the competing methods for image-level lesion detection using synthetic MRI images. In terms of the area under the curve (AUC), Fixed-Point GAN achieves comparable performance with ResNet-50 classifier, but substantially outperforms StarGAN, f-AnoGAN, and Alex, \etal. Note that, for f-AnoGAN, we use the average activation of difference maps as the detection score, because we find it more effective than using the maximum activation of difference maps and also more effective than the anomaly scores proposed in the original work. 

\figurename~\ref{subfig:brats_froc} shows the Free-Response ROC (FROC) analysis for  synthetic MR images. Our Fixed-Point GAN achieves a sensitivity of 84.5\% at 1 false positive per image, outperforming StarGAN, f-AnoGAN, and Alex, \etal with the sensitivity levels of 13.6\%, 34.6\%, 41.3\% at the same level of false positive. The ResNet-50-CAM at 32x32 resolution achieves the best sensitivity level of 60\% at 0.037 false positives per image. 
Furthermore, we compare ResNet-50-CAM with Fixed-Point GAN using mean IoU (intersection over union) score, obtaining mean IoU of 0.2609$\pm$0.1283 and  0.3483$\pm$0.2420, respectively. Similarly, ROC and FROC analysis on real MRI images are provided in \figurename~\ref{subfig:brats_real_roc} and \figurename~\ref{subfig:brats_real_froc}, respectively, showing that our method is outperformed at the low false positive range, but achieves a significantly higher sensitivity overall. Qualitative comparisons between StarGAN, Fixed-Point GAN, CAM, and f-AnoGAN for brain lesion detection and localization are provided in \figurename~\ref{fig:detection_localization_results}. More qualitative comparisons are available in the appendix.

\subsection{Pulmonary Embolism Detection and Localization with Image-Level Annotation}
\label{subsec:application_pe}

\noindent\textbf{Dataset.} Pulmonary embolism (PE) is a blood clot that travels from a lower extremity source to the lung, where it causes blockage of the pulmonary arteries. It is a major national health problem, but computer-aided PE detection and localization can improve diagnostic capabilities of radiologists for the detection of this disorder, leading to earlier and effective therapy for this potentially deadly disorder. We utilize a database consisting of 121 computed tomography pulmonary angiography (CTPA) scans with a total of 326 emboli. The dataset is pre-processed as suggested in~\cite{zhou2017fine,tajbakhsh2016convolutional,tajbakhsh2019computer}, divided at the patient-level into a training set with 3,840 images, and a test set with 2,415 images. Further details are provided in the appendix.

\noindent\textbf{Method and Evaluation.} As with brain lesion detection and localization (\sectionname~\ref{subsec:application_brain}), we use only image-level annotations during training. At inference time, we always remove PE from the input image (\ie\ translating both PE and non-PE images into the non-PE domain) irrespective of whether PE is present or absent in the input image. We follow the same procedure described in \sectionname~\ref{subsec:application_brain} to generate the difference maps, detection scores, and ROC curves. Note that, since each PE image has an embolus in its center, an embolus is considered as ``detected" if the corresponding PE image is correctly classified; otherwise, the embolus is considered ``missed". As such, unlike \sectionname~\ref{subsec:application_brain}, we do not pursue a connected component analysis for PE localization.

We compare our Fixed-Point GAN with StarGAN and ResNet-50. We have excluded GAN-based method~\cite{alex2017generative} and f-AnoGAN from the quantitative comparisons because, despite our numerous attempts, the former encountered convergence issues and the latter produced poor detection and localization performance. 
Nevertheless, we have provided images generated by f-AnoGAN in appendix.

\noindent\textbf{Results.} \figurename~\ref{subfig:pe_roc} shows the ROC curves for image-level PE detection. Fixed-Point GAN achieves an AUC of 0.9668 while StarGAN and ResNet-50 achieve AUC scores of 0.8832 and 0.8879, respectively. 
\figurename~\ref{subfig:pe_froc} shows FROC curves for PE localization.
Fixed-Point GAN achieves a sensitivity of 97.2\% at 1 false positive per volume, outperforming StarGAN and ResNet-50 with sensitivity levels of of 88.9\% and 80.6\% at the same level of false positives per volume. The qualitative comparisons for PE removal between StarGAN and Fixed-Point GAN are given in \figurename~\ref{fig:detection_localization_results}.

\begin{figure}[t]
     \centering
     \begin{subfigure}[b]{0.445\linewidth}
         \centering
         \caption{Image-level detection}
         \includegraphics[width=\linewidth]{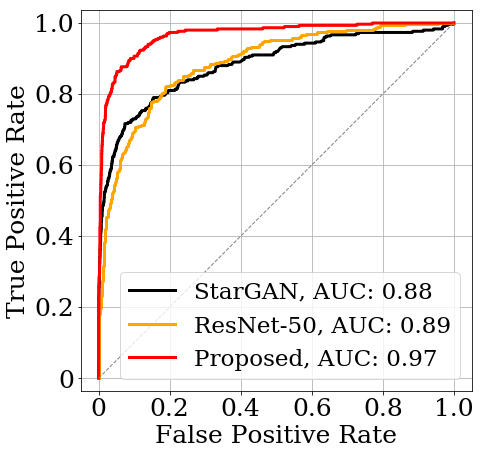}
         \label{subfig:pe_roc}
     \end{subfigure}
     \begin{subfigure}[b]{0.535\linewidth}
         \centering
         \caption{Lesion-level localization}
         \includegraphics[width=\linewidth]{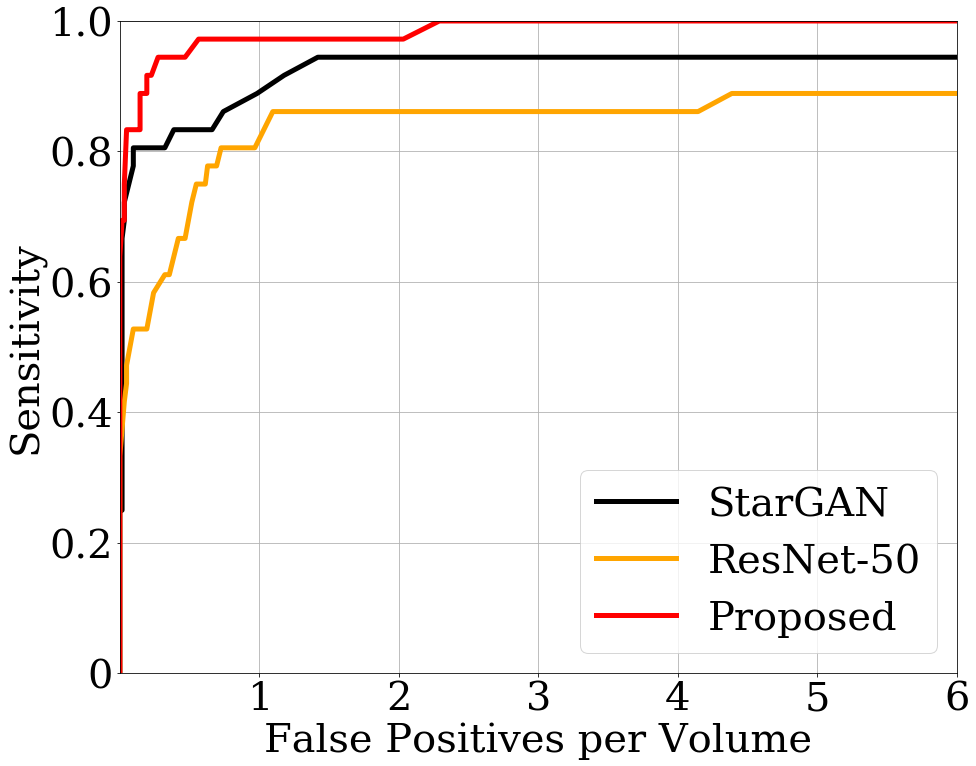}
         \label{subfig:pe_froc}
     \end{subfigure}
     \vspace{-6pt}
     \caption{Comparing Fixed-Point GAN with StarGAN, f-AnoGAN, and ResNet-50 on the PE dataset. (a) ROCs for image-level detection. (b) FROCs for lesion-level localization.}
     \label{fig:pe_result}
     \vspace{-6pt}
\end{figure}

\subsection{Discussions}
\label{sec:discussion}

In \figurename~\ref{fig:brats_result}, we show that StarGAN performs poorly for image-level brain lesion detection, because StarGAN is designed to perform general-purpose image translations, rather than an image translation suitable for the task of disease detection. Owing to our new training scheme, Fixed-Point GAN can achieve precise image-level detection.

Comparing \figurename~\ref{fig:brats_result} and~\ref{fig:pe_result}, we observe that StarGAN performs far better for PE than brain lesion detection. We believe this is because brain lesions can appear anywhere in the input images, whereas PE always appears in the center of the input images, resulting in a less challenging problem for StarGAN to solve. Nonetheless, Fixed-Point GAN outperforms StarGAN for PE detection, achieving an AUC score of 0.9668 compared to 0.8832 by StarGAN.

Referring to \figurename~\ref{fig:detection_localization_results}, we further observe that neither StarGAN nor Fixed-Point GAN can completely remove large objects, like sunglasses or brain lesions, from the images. Nevertheless, for image-level detection and lesion-level localization, it is sufficient to remove the objects partially, but precise lesion-level segmentation using an image-to-image translation network requires complete removal of the object. This challenge is the focus for our future work.

\section{Conclusion}
\label{sec:conclusion}

We have introduced a new concept called fixed-point translation, and developed a new GAN called Fixed-Point GAN. Our comprehensive evaluation demonstrates that our Fixed-Point GAN outperforms the state of the art in image-to-image translation and is significantly superior to predominant anomaly detection and weakly-supervised localization methods in both disease detection and localization with only image-level annotation. The superior performance of Fixed-Point GAN is attributed to our new training scheme, realized by supervising same-domain translation and regularizing cross-domain translation.

\medskip
\noindent {\bf Acknowledgments:} This research has been supported partially by ASU and Mayo Clinic through a Seed Grant and an Innovation Grant, and partially by NIH under Award Number R01HL128785. The content is solely the responsibility of the authors and does not necessarily represent the official views of NIH. We thank Zuwei Guo for helping us with the implementation of a baseline method.

\newpage
{\small
\bibliographystyle{ieee_fullname}
\bibliography{iccv2019}
}

\newpage
\onecolumn
\appendix
\section*{Appendix}

\jlreplace{
}{
}
\begin{figure}[h]
    \centering 
    \includegraphics[width=0.8\linewidth]{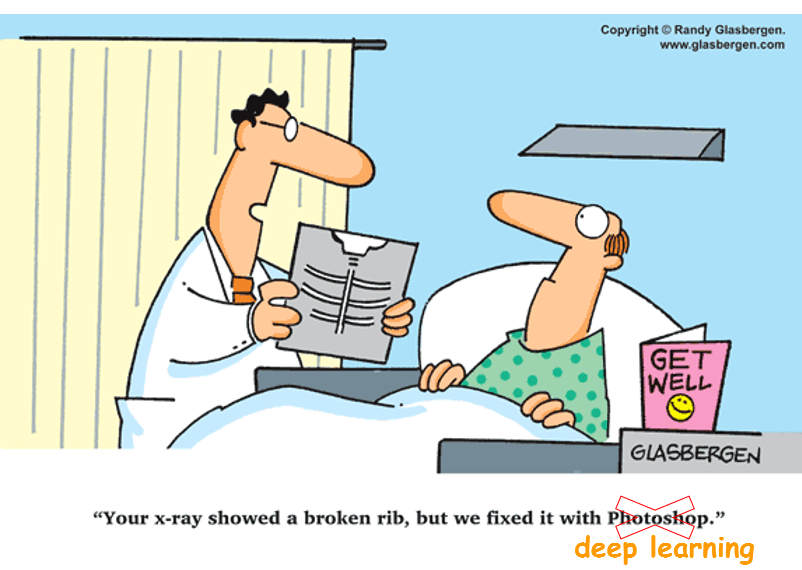}
   \caption{Now, all humor aside, if a GAN can remove diseases {\em completely} from images, it will offer an {\em ideal} method for segmenting {\em all} diseases, an ambitious goal that has yet to be achieved. Nevertheless, our method is still of great clinical significance in computer-aided diagnosis for medical imaging, because it can be used for disease detection and localization by subtracting the generated healthy image from the original image to reveal diseased regions, that is, {\bf detection and localization by removal}. More importantly,  our Fixed-Point GAN is trained using {\em only} image-level annotation. It is much easier to obtain image-level annotation than lesion-level annotation, because a large number of diseased and healthy images can be collected from PACS (picture archiving and communication systems), and labeled at the image level by analyzing their radiological reports through NLP (natural language processing). With the availability of large {\em well-organized} databases of medical images and their corresponding radiological reports in the future, we envision that  Fixed-Point GAN will be able to detect and localize diseases more accurately---and eventually to segment diseases---using only image-level annotation. [The cartoon was provided courtesy of Karen Glasbergen with permission for adaptation and modification]}
    \label{fig:joke}
\end{figure}

\vfill
\begin{center}
{\sf All figures and images} ({\small including those in the main paper}) {\sf better viewed on-line in color and magnified for details}
\end{center}
\vfill

\newpage
\subsection*{Eyeglass Detection and Localization by Removal Using Only Image-Level Annotation of the CelebA Dataset}

\begin{figure}[h]
    \centering 
    \includegraphics[width=1.0\linewidth]{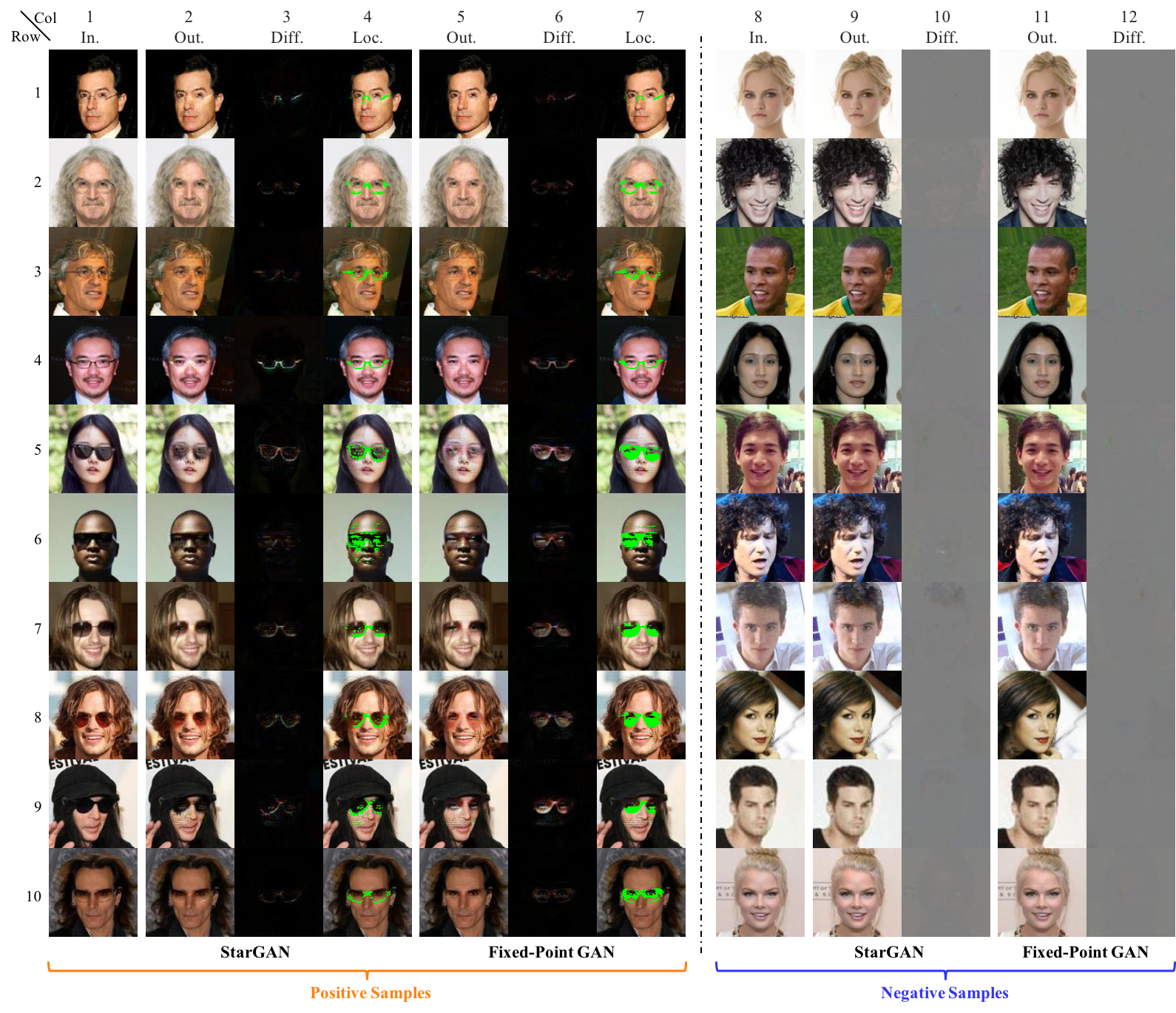}
    \vspace{-3pt}
    \caption{[Continued from \figurename~\ref{fig:detection_localization_results}] Additional test results in eyeglass detection and localization by removal. The difference map (Column 3 for StarGan; Column 6 for Fixed-Point GAN) shows the absolute difference between the input (Column 1) and output (Column 2 for StarGAN; Column 5 for Fixed-Point GAN). Applying the $k$-means clustering algorithm on the difference map yields a localization map, which is then superimposed on the original image (Column 4 for StarGAN; Column 7 for Fixed-Point GAN), showing both StarGAN and Fixed-Point GAN attempt to remove eyeglasses. However, the former leaves noticeable white ``inks'' along eyeglass frames (Rows 1 and 4, Column 2), while our method better preserves the face color. Removing sunglasses (Rows 5--9) has proven to be challenging: both methods suffer from partial removal and artifacts. Nevertheless, Fixed-Point GAN tends to recover the face under the glasses and frames, but StarGAN only changes regions around the frames. More importantly, our method can ``insert" eyes at proper positions, as revealed in the difference maps (Rows 5--9, Column 6), while StarGAN can hardly do so. 
    To better visualize the subtle changes for negative samples (Column 8), instead of the absolute difference, we show the difference directly, where the gray color (\ie, 0) means ``no change". In this way, it can be observed more easily that StarGAN does some unnecessary small changes on hair (Rows 7 and 9, Column 10) and eyes (Rows 7 and 10, Column 10), while Fixed-Point GAN generates smooth gray images (\ie., close to 0 everywhere; Column 12). Please note that the CelebA Dataset currently does not have ground truth on the location and segmentation of glasses; therefore, a quantitative performance evaluation of eyeglass localization cannot be conducted. However, our quantitative performance evaluations of brain lesion localization and pulmonary embolism localization are included in \sectionname~\ref{sec:applications}.}    
    \label{fig:eyeglasses_removal}
\end{figure}

\newpage
\subsection*{Multi-Domain Image-to-Image Translation}

\begin{figure}[h]
    \centering 
    \includegraphics[width=1.0\linewidth]{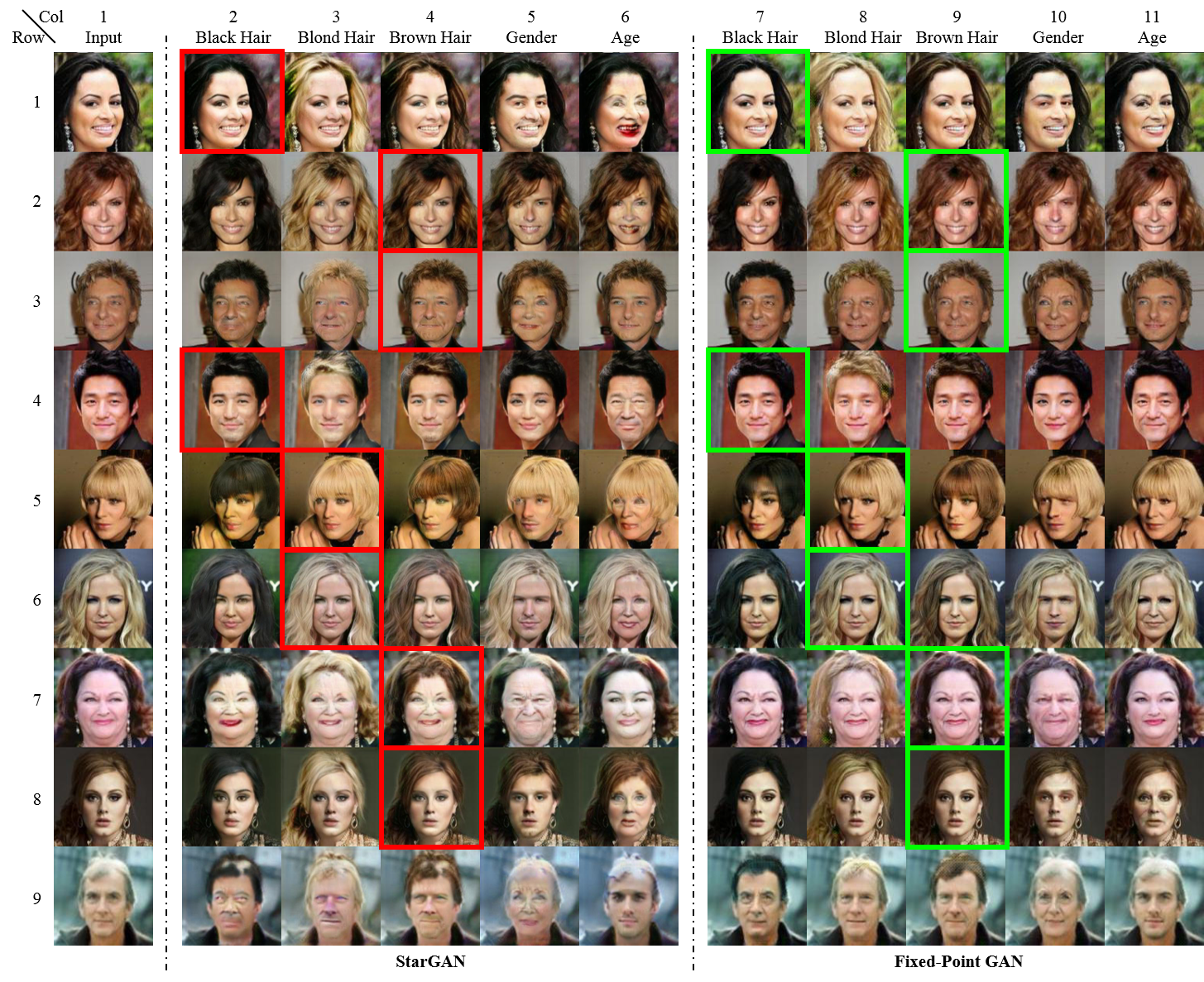}
    \vspace{-3pt}
    \caption{[Continued from \figurename~\ref{fig:celeba5_stargan_vs_ours}] More test results in multi-domain image-to-image translation on CelebA dataset. Visually, Fixed-Point GAN outperforms StarGAN: Fixed-Point GAN (Columns 7-11) better preserves the background (Rows 1, 3, 4, 6, 8, and 9), face color (Rows 2--7), and facial features (Rows 7 and 9), whereas StarGAN (Columns 2-6) makes unnecessary changes. Furthermore, for same-domain translation, StarGAN introduces noticeable artifacts (outlined in red), while Fixed-Point GAN can leave all the details intact (outlined in green). It is worthy noting that the hair color of the facial image in the last input row (\ie, Row 9, Column 1) belongs to Domain \texttt{gray hair}, which is not included in the training phase. As can be seen, Fixed-Point GAN successfully translates the input image to target domains by changing the unseen hair color to desired colors and maintaining the original hair color (gray) in hair-color-unrelated translations (Row 9, Columns 10--11). However, StarGAN produces unnatural images with artifacts (Row 9, Columns 2--4) and inconsistent white hair colors (Row 9, Columns 5--6). This example shows that Fixed-Point GAN outperforms StarGAN in generalization.}    
    \label{fig:multi_domain_translation}
\end{figure}

\newpage
\subsection*{Brain Lesion Detection and Localization by Removal Using Only Image-Level Annotation}

\begin{figure}[h]
    \centering 
    \includegraphics[width=1.0\linewidth]{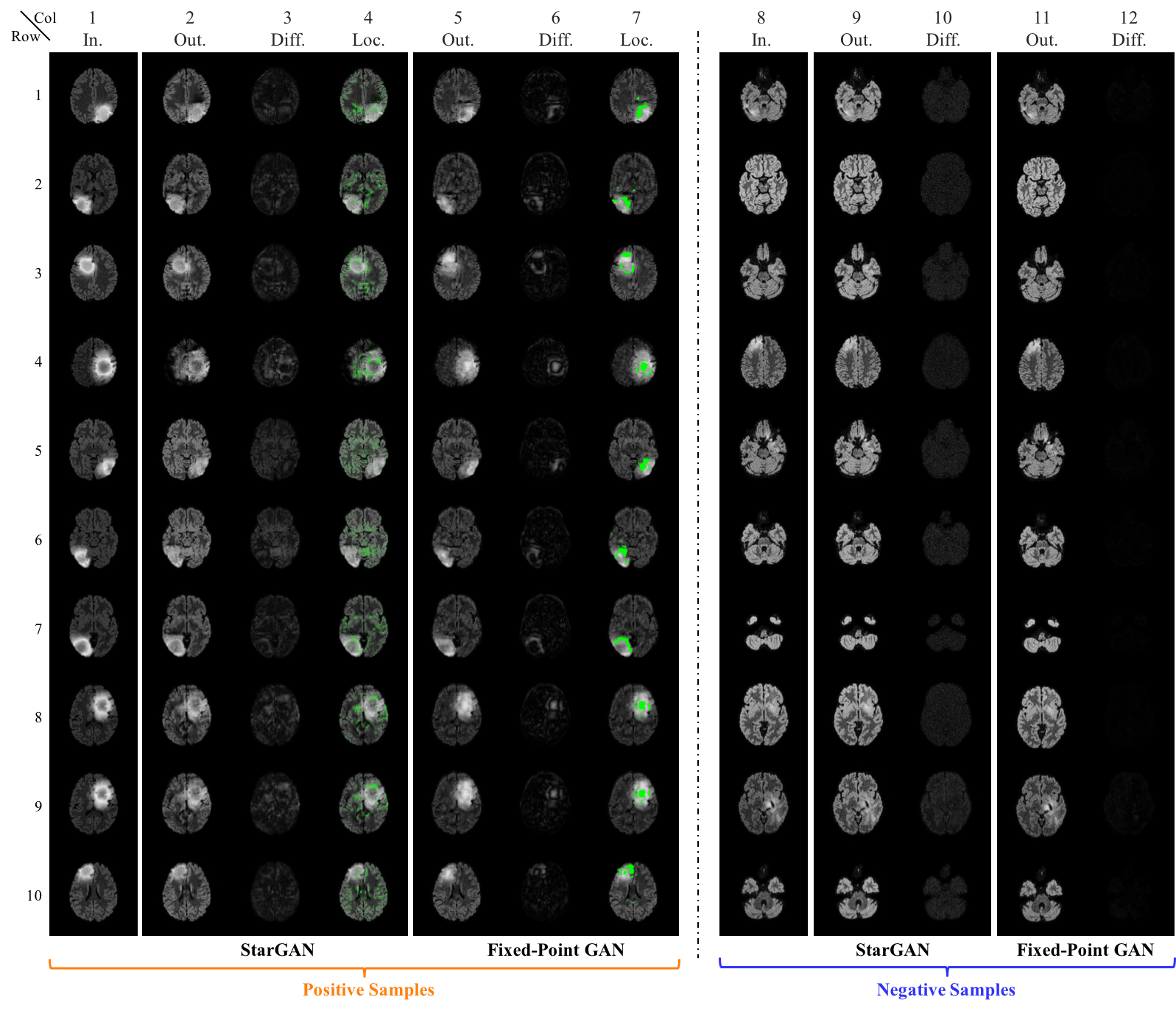}
    \vspace{-3pt}
    \caption{[Continued from \figurename~\ref{fig:detection_localization_results}] Brain lesion detection and localization tested on additional positive samples (\ie brain images with lesions; Column 1) and negative samples (\ie brain images without lesions; Column 8). Fixed-Point GAN achieves superior detection performance, benefiting from the cleaner difference maps of negative samples (Column 12), while StarGAN highlights the brain regions in all cases, thereby rendering the difference maps of positive and negative samples indistinguishable (comparing Column 3 with Column 10). Although both methods fail to remove lesions completely, our method focuses on the lesion regions, and consequently, it produces higher localization accuracy. In contrast, the StarGAN localization map (Column 4) is very noisy and unsuitable for lesion localization. These comparisons demonstrate the superiority of Fixed-Point GAN in lesion detection and localization. For quantitative performance evaluations, please refer to \figurename~\ref{fig:brats_result} and \sectionname~\ref{subsec:application_brain}.}
    \label{fig:brats_positives}
\end{figure}

\newpage
\subsection*{Pulmonary Embolism Detection and Localization by Removal Using Only Image-Level Annotation}

\begin{figure}[h]
    \centering 
    \includegraphics[width=1.0\linewidth]{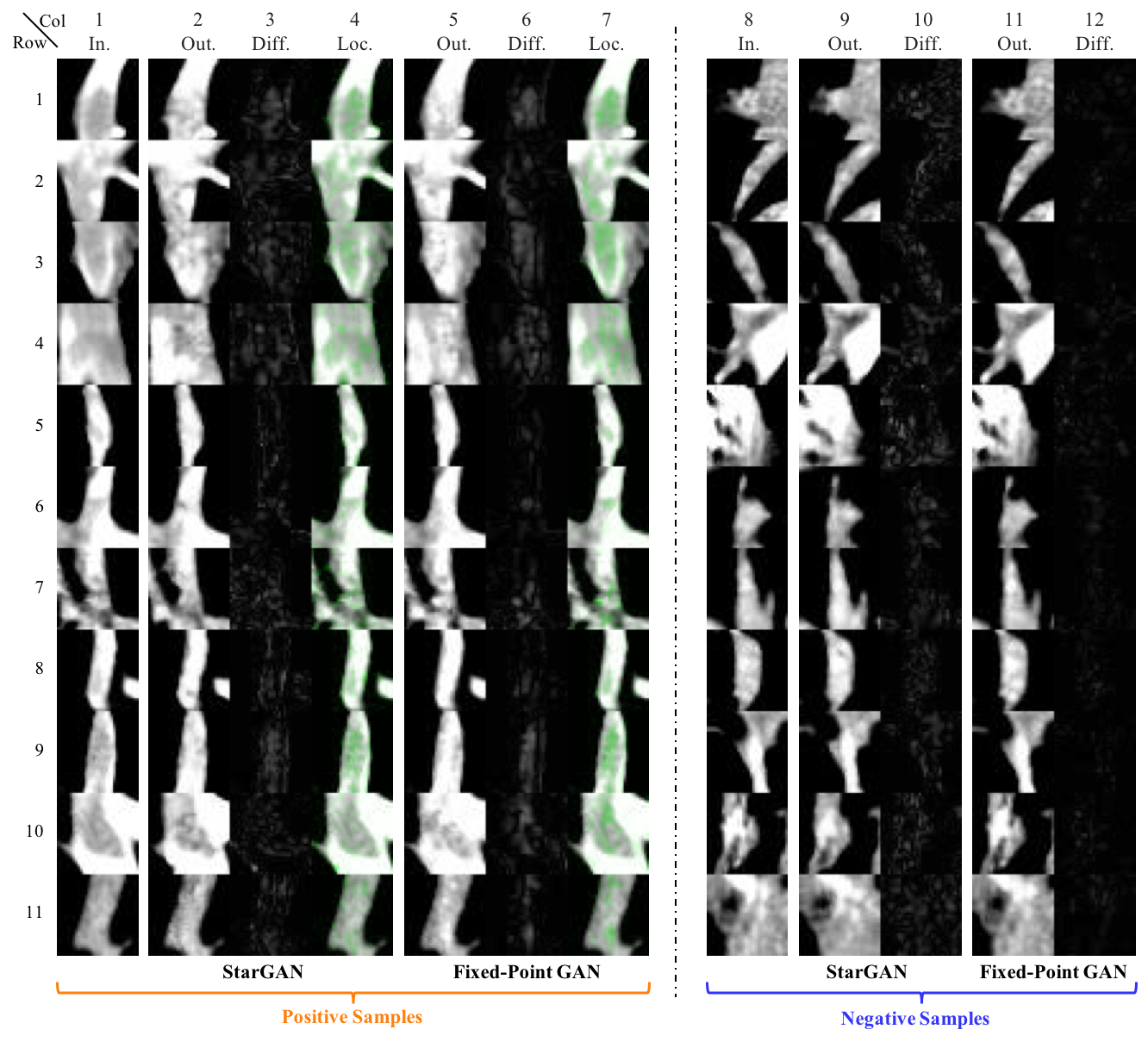}
    \vspace{-3pt}
    \caption{[Continued from \figurename~\ref{fig:detection_localization_results}] Pulmonary Embolism (PE) detection and localization (longitudinal view) tested on additional positive samples (\ie images with PEs; Column 1) and negative samples (\ie images without PEs; Column 8). PE is a blood clot that creates blockage (appearing dark and centered within the image) in pulmonary arteries (which appear white). The current candidate generator (\eg \cite{liang2007computer}) produces many false positive results (negative samples) during localization; therefore, our goal in this application is to reduce false positives through StarGAN and Fixed-Point GAN.
    Compared with StarGAN, the difference maps of negative samples from Fixed-Point GAN is clean and easy to be separated from the difference maps of positive samples, yielding better detection performance. For quantitative performance evaluations, please refer to \figurename~\ref{fig:pe_result} and \sectionname~\ref{subsec:application_pe}.}
    \label{fig:pe_1_supp}
\end{figure}

\newpage
\begin{figure}[h]
    \centering 
    \includegraphics[width=1.0\linewidth]{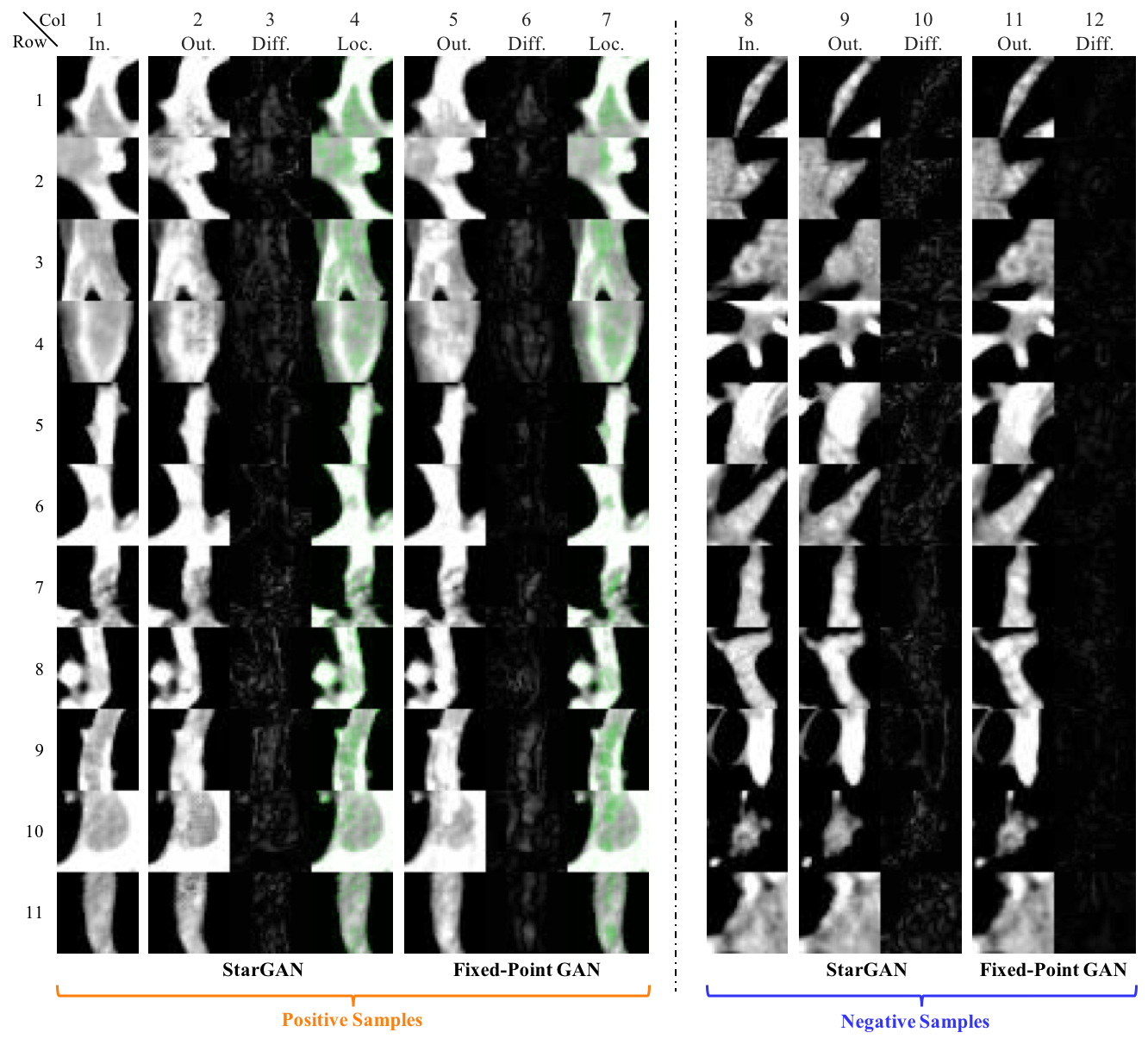}
    \vspace{-3pt}
    \caption{Pulmonary Embolism (PE) detection and localization (longitudinal view). Notice the images are from the same candidates as \figurename~\ref{fig:pe_1_supp} but the view direction is orthogonal to the angle used in \figurename~\ref{fig:pe_1_supp}.}
    \label{fig:pe_2_supp}
\end{figure}

\newpage
\begin{figure}[h]
    \centering 
    \includegraphics[width=1.0\linewidth]{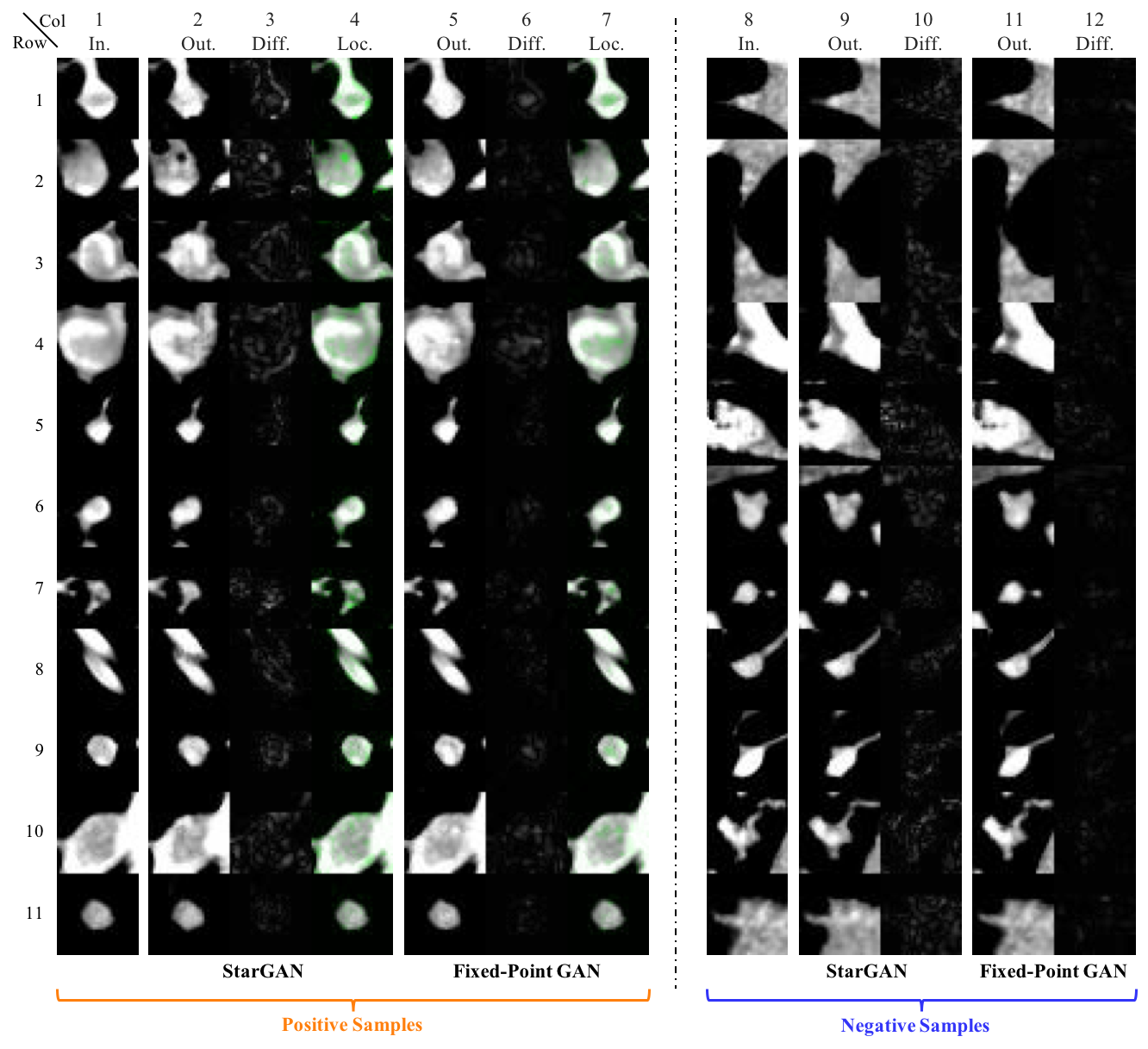}
    \vspace{-3pt}
    \caption{Pulmonary Embolism (PE) detection and localization (cross-sectional view). Notice the images are from the same candidates as \figurename~\ref{fig:pe_1_supp} but the orientation is cross-sectional.}
    \label{fig:pe_3_supp}
\end{figure}

\newpage
\subsection*{Localization Using Class Activation Maps (CAMs)}

\begin{figure}[h]
    \centering 
    \includegraphics[width=1.0\linewidth]{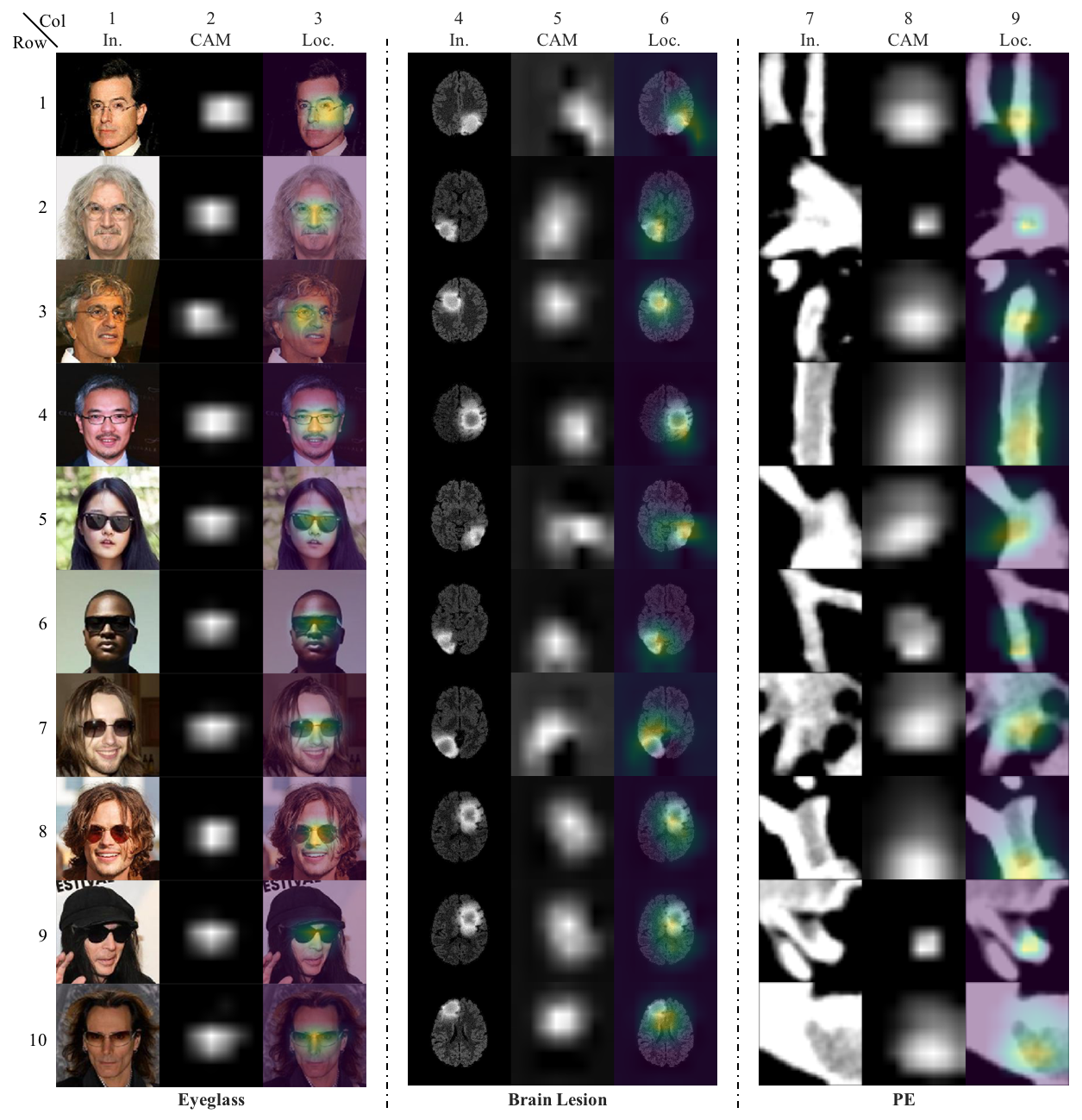}
    \vspace{-3pt}
    \caption{[Continued from \figurename~\ref{fig:detection_localization_results}] Additional test results of localization using class activation maps (CAMs). CAMs for localizing glasses, brain lesion, and PE are obtained from ResNet-50 classifiers trained with corresponding datasets. Localization using CAMs is not as precise as Fixed-Point GAN, as discussed in \sectionname~\ref{subsec:application_brain}.}
\end{figure}

\newpage
\subsection*{Qualitative Results of f-AnoGAN and VA-GAN for Brain Lesion Detection}

\begin{figure}[h]
    \centering 
    \includegraphics[width=1.0\linewidth]{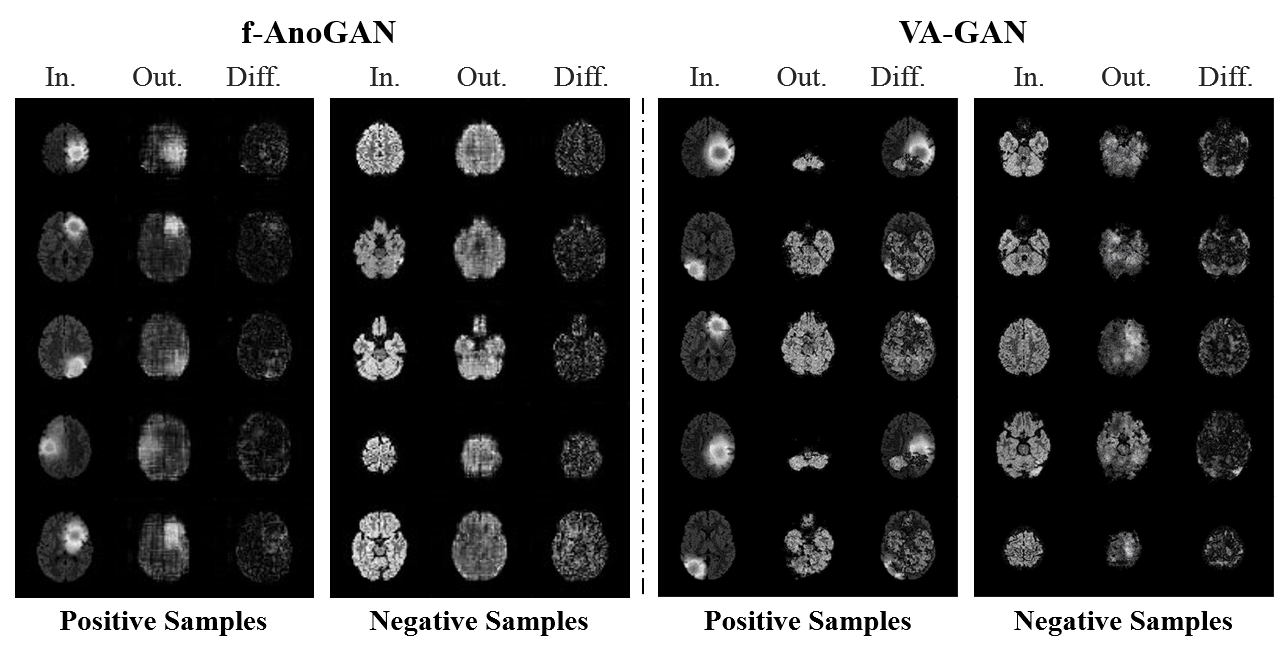}
    \vspace{-3pt}
    \caption{[Continued from \figurename~\ref{fig:detection_localization_results}] In spite of learning healthy images only, f-AnoGAN~\cite{schlegl2019f} performs competitively on image-level detection task (see \figurename~\ref{subfig:brats_roc}), however, noisy difference maps impede its localization power (see \figurename~\ref{subfig:brats_froc}). On the other hand, VA-GAN~\cite{baumgartner2017visual} fails to preserve anatomical structures when trained with unpaired images, thus violates our Req.~\ref{req2} and renders unsuitable for our purpose.}
\end{figure}

\subsection*{Qualitative Results of f-AnoGAN on the PE Dataset}

\begin{figure}[h]
    \centering 
    \includegraphics[width=1.0\linewidth]{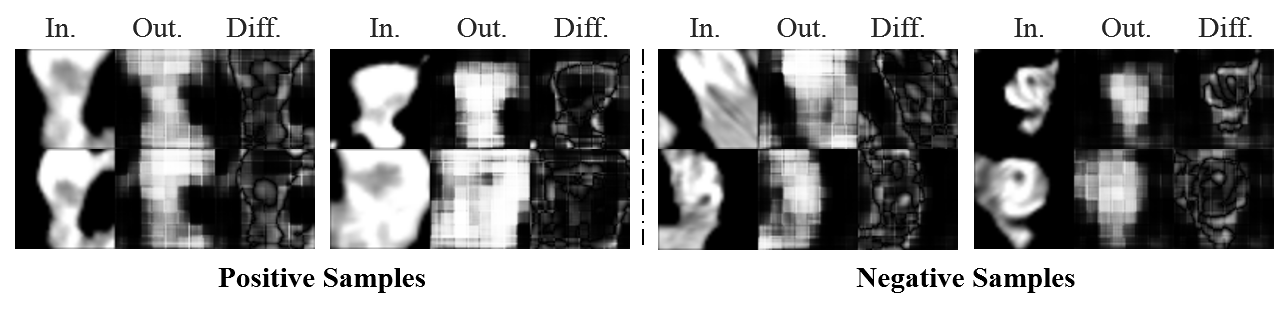}
    \vspace{-3pt}
    \caption{As discussed in \sectionname~\ref{subsec:application_pe}, f-AnoGAN~\cite{schlegl2019f} fails to produce good quality images for the PE dataset, resulting in noisy difference maps. Hence, performs miserably on detection and localization tasks for the PE dataset.}
\end{figure}

\newpage

\subsection*{Dataset Processing Details}
\label{sec:dataset_appending}

\noindent\textbf{Brain Lesion Detection and Localization with Image-Level Annotation:} BRATS 2013 consists of synthetic and real images, where each of them is further divided into high-grade gliomas (HG) and low-grade gliomas (LG). There are 25 patients with both synthetic HG and LG images and 20 patients with real HG and 10 patients with real LG images. For each patient, \texttt{FLAIR}, \texttt{T1}, \texttt{T2}, and \texttt{post-Gadolinium T1} magnetic resonance (MR) image sequences are available. To ease the analysis, we keep the input features consistent by using only one MR imaging sequence (FLAIR) for all patients in both the HG and LG categories, resulting in a total of 9,050 synthetic MR slices and 5,633 real MR slices. We further pre-process the dataset by removing all slices that are either blank or have very little brain information. Finally, we randomly select 40 patients with 5,827 slices for training and 10 patients with 1,461 slices for testing from synthetic MRI images. For the experiments on real MRI images, we randomly select 24 patients with 3,044 slices for training and 6 patients with 418 slices for testing. During training, we set aside one batch of the random samples from the training dataset for validation. We pad the slices to 300$\times$300 and then center-crop to 256$\times$256, ensuring that the brain regions appear in the center of the images. Each pixel in the dataset is assigned one of the five possible labels: 1 for non-brain, non-tumor, necrosis, cyst, hemorrhage; 2 for surrounding edema; 3 for non-enhancing tumor; 4 for enhancing tumor core; and 0 for everything else. We assign an MR slice to the healthy domain if all contained pixels are labeled as 0; otherwise, the MR slice is assigned to the diseased domain.

\medskip

\noindent\textbf{Pulmonary Embolism Detection and Localiza-tion with Image-Level Annotation:} We utilize a database consisting of 121 computed tomography pulmonary angiography (CTPA) scans with a total of 326 emboli. The dataset is pre-processed as suggested in~\cite{zhou2017fine,tajbakhsh2016convolutional,tajbakhsh2019computer}. A candidate generator~\cite{liang2007computer} is first applied to generate a set of PE candidates, and then by comparing against the ground truth, the PE candidates are labeled as PE or non-PE. Finally, a 2D patch of size 15$\times$15mm is extracted around each PE candidate according to a vessel-aligned image representation~\cite{tajbakhsh2015computer}. As a result, PE appears at the center of the PE images. The extracted images are rescaled to 128$\times$128. The dataset is divided at the patient-level into a training set with 434 PE images (199 unique PEs) and 3,406 non-PE images, and a test set with 253 PE images (127 unique PEs) and 2,162 non-PE images. To enrich the training set, rotation-based data augmentation is applied for both PE and non-PE images.

\newpage

\subsection*{Implementation Details}
\label{sec:implementation}

\begin{table*}[h]
    \begin{center}
        \setlength\tabcolsep{3pt}
        \begin{tabular}{l c c c c | c c c}
            & \multicolumn{4}{c |}{\textbf{Image-Level Detection (AUC)}} & \multicolumn{3}{c}{\textbf{Lesion-Level Loc. Sensitivity at 1 False Positive}} \\
            \hline
            \textbf{Dataset} & \textbf{StarGAN} & \textbf{w/ Delta} & \textbf{w/ Fixed-Point Translation} & \textbf{w/ Both} & \textbf{StarGAN} & \textbf{w/ Fixed-Point Translation} & \textbf{w/ Both} \\
            \hline
            BRATS & 0.4611 & 0.5246 & 0.9980 & \textbf{0.9831} & 13.6\% & 81.2\% & \textbf{84.5\%} \\
            PE & 0.8832 & 0.8603 & 0.9216 & \textbf{0.9668} & 88.9\% & 94.4\% & \textbf{ 97.2\%} \\
            \hline
        \end{tabular}
    \end{center}

    \caption{Ablation study of the generator's configuration on brain lesion (BRATS 2013) and pulmonary embolism (PE) detection. Selected combinations are in bold. The columns ``w/Delta'', ``w/Fixed-Point Translation'', and ``w/Both'' mean StarGAN trained with only delta map, only fixed-point translation learning, and both of them combined, respectively. The empirical results show that the performance gain is largely due to fixed-point translation learning---the contribution by the delta map is minor and application-dependent.}
    \label{tab:ablation_study}
\end{table*}

We have revised adversarial loss (Eq.~\ref{eq:adv_loss}) based on the Wasserstein GAN~\cite{arjovsky2017wasserstein} objective by adding a gradient penalty~\cite{gulrajani2017improved} to stabilize the training, which is defined as

\begin{equation}
\begin{split}
\mathcal{L}_{adv} = \thinspace & {\mathbb{E}}_{x}[{D}_{real/fake}(x)] - \sum_{c \in \{c_x, c_y\}}{\mathbb{E}}_{x, c}[{D}_{real/fake}(G(x, c))] - {\lambda}_{gp} \thinspace {\mathbb{E}}_{\hat{x}}[{{(||{\triangledown}_{\hat{x}} {D}_{real/fake}(\hat{x})||}_{2} - 1)}^{2}] \thinspace,
\end{split}
\end{equation}

Here, $\hat{x}$ is uniformly sampled along a straight line between a pair of a real and a fake image. The gradient penalty coefficient (${\lambda}_{gp}$) is set to 10 for all experiments. Values for $\lambda_{domain}$ and $\lambda_{cyc}$, are set at 1 and 10, respectively, for all experiments. $\lambda_{id}$ is set to 10 for CelebA, 0.1 for BRATS 2013, and 1 for PE dataset. 200K iteration is found to be sufficient for CelebA and the PE dataset, whereas BRATS 2013 requires 300K iteration for generating good quality images. To facilitate a fair comparison, we use the same generator and discriminator architectures as the public implementation of StarGAN. All models are trained using the Adam optimizer with learning rate $1e^{-4}$ for both the generator and discriminator across all experiments.

Following~\cite{baumgartner2017visual}, we slightly change the architecture of the generator to predict a residual (delta) map rather than the desired image directly. Specifically, the generator's output is computed by adding the delta map to the input image, followed by the application of a $tanh$ activation function, $tanh(G(x,c)+x)$. Our ablation study, summarized in \tablename~\ref{tab:ablation_study}, shows the disease detection and localization performance of StarGAN (baseline approach), and the incremental performance improvement using delta map learning, fixed-point translation learning, and the two approaches combined. We find that the major improvement over StarGAN comes from fixed-point translation learning, but the combined approach, for most cases, provides enhanced performance compared to each individual approach (see \tablename~\ref{tab:ablation_study}).  We therefore use the combination of delta map learning and fixed-point translation learning in our proposed Fixed-Point GAN, noting that the major improvement over StarGAN is due to the proposed fixed-point translation learning scheme. The implementation is publicly available at \url{http://github.com/jlianglab/Fixed-Point-GAN}

\end{document}